\documentclass{aa}

\usepackage{graphicx}
\newcommand{\km}{\,\mbox{km}\,\mbox{s}^{-1}}
\def\Ha{\hbox{H$\alpha$~}}
\def\N{\symbol{242}}

\begin{document}

\title{The  orientation parameters and rotation curves of
15 spiral galaxies.
\thanks{Based on observations collected with the 6m telescope of the
Special Astrophysical Observatory (SAO) of the Russian Academy of
Sciences (RAS), operated under the financial support of the
Science Department of Russia (registration number 01-43).} }

\titlerunning{The orientation parameters of spiral galaxies.}

\authorrunning{Fridman et al.}

\author{A.M.~Fridman$^{1,\,2}$, V.L.~Afanasiev$^3$, S.N.~Dodonov$^3$,
O.V.~Khoruzhii$^{1,\,4}$,  A.V.~Moiseev$^{3,\,5}$,
O.K.~Sil'chenko$^{2,\,6}$, \and A.V.~Zasov$^2$. }

\institute{ Institute of Astronomy of the Russian Academy of
Science, 48, Pyatnitskaya St., Moscow, 109017, Russia \and
Sternberg Astronomical Institute, Moscow State University,
University prospect, 13, Moscow, 119992, Russia \and Special
Astrophysical Observatory, Nizhnij Arkhyz, Karachaevo- Cherkesia,
357147,  Russia, \and Troitsk Institute for Innovation and
Thermonuclear Researches, Troitsk, Moscow reg., 142092, Russia
\and Guest investigator of the UK Astronomy Data Centre
\and Isaac Newton Institute of Chile, Moscow Branch}
\date{Received ...../ Accepted .....}

\abstract{We analyzed ionized gas motion and disk orientation
parameters for 15 spiral galaxies. Their velocity fields were
measured with the H$\alpha$ emission line by using the Fabry-Perot
interferometer at the 6m telescope of SAO RAS. Special attention
is paid to the problem of estimating the position angle of the
major axis $(PA_0)$ and the inclination ($i$) of a disk, which
strongly affect the derived circular rotation velocity. We
discuss and compare different methods of obtaining these
parameters from kinematic and photometric observations, taking
into account the presence of regular velocity (brightness)
perturbations caused by spiral density waves. It is shown that
the commonly used method of tilted rings may lead to systematic
errors in the estimation of orientation parameters (and hence of
circular velocity) being applied to galaxies with an ordered
spiral structure. Instead we recommend using an assumption of
constancy of $i$ and $PA_0$ along a radius, to estimate these
parameters. For each galaxy of our sample we present
monochromatic H$\alpha$- and continuum maps, velocity fields of
ionized gas, and the mean rotation curves in the frame of a model
of pure circular gas motion. Significant deviations from circular
motion with amplitudes of several tens of $\km$ (or higher) are
found in almost all galaxies. The character and possible nature of
the non-circular motion are briefly discussed.
\bigskip
}

\maketitle

\section{Introduction}

Most of the available data on the kinematics of ionized gas in
disk galaxies is obtained with long-slit spectrographs, allowing
us to construct mean rotation curves under an assumption of pure
circular rotation.  But the circular rotation model, being useful
for determinations of the global properties of a galaxy, is only a
zeroth order approximation to the real and rather complex
dynamical picture.

Even if there is no active nucleus or tidal perturbation from a
nearby companion to affect gas kinematics in a galaxy, the
observed non-circular motion may have amplitudes up to $30-50
\km$ (or even higher in some cases), being therefore supersonic.
The nature of non-circular motion may be different. Small-scale
gas motion is usually related to star formation sites (HII
complexes), whereas large-scale velocity perturbations are caused
by wave spiral pattern and/or central oval (bar) impacts on the
gaseous disk.

At the end of the 80s, a series of observations with a long-slit
spectrograph was undertaken at the 6m telescope of the SAO RAS,
which have revealed significant non-circular motion with
amplitudes of $30-100 \km$ in ordinary spiral galaxies (Afanasiev
et al.  1988 -- 1992).  But the most complete information on the
gas kinematics in galactic disks can be acquired only by using
full 2D velocity fields which may be obtained in the optical
spectral range by methods of panoramic (2D) spectroscopy. The
method based on the Fabry-Perot interferometric observations was
used by several groups of observers to get the velocity fields
for large samples of different morphological types of galaxies
(see f.e. Schommer et al., 1993, Amram et al. 1995). The largest
project of this type is the GHASP survey, carried out at the
Haute-Provence Observatory, which should provide a homogeneous
sample of 2D velocity fields of about 200 spiral and irregular
galaxies. This survey aimes to compare galaxies in different
environments, different stages of evolution and at different
redshifts (Garrido et al. 2002, 2003). Unlike the works mentioned
above, the main goal of our project is to study the regular
deviations of gas motion from circular rotation in spiral
galaxies, which may be caused by spiral density waves -- in
parallel with the routine task of obtaining the rotation curves
and mass distributions. For this purpose we use the Scanning
Fabry-Perot interferometer (IFP) of the 6m telescope. We have
obtained high-resolution line-of-sight velocity fields of the
ionized gas over a large field of view (about $3'$) for 42 spiral
galaxies,\footnote{For the full list, see \par
http://www.sao.ru/hq/lsfvo/vortex/vortex.html} both barred and
non-barred, mostly of Sb-Sc types, by measuring the H$\alpha$
(and in some cases also [NII]) emission line. This sample may be
by no means considered as a complete one to some limiting
magnitude or angular size. It just represents spiral galaxies of
different types with the angular diameter of $2' < D_{25} < 6'$
(for the most of them) having various types of spiral patterns --
from certain grand-design to flocculent ones.  The results of the
velocity field analysis for some of the galaxies of our sample
have already been published (Fridman et al. 1997, 1998, 1999,
2001a,b,c; Fridman \& Khoruzhii, 2003). In this work we present
the results of observations and the rotation curves for 15
galaxies of the sample. In parallel with the Doppler velocities
measurements, brightness maps in H$\alpha$ and near-H$\alpha$
continuum were also obtained for each galaxy.

In principle, the Doppler velocity field contains information
both on the rotation and on non-circular gas motion, including
the ordered wave motion related to a spiral density wave (if it
exists in a chosen galaxy). The problem to be solved is the
development of reliable methods which would enable us to extract
and analyze these motions separately.  Historically, the first
attempts at such an approach were based on drawing a ``mean''
rotation curve, and on the consequent extraction of non-circular
motion as residuals left after the subtraction of the circular
rotation component from the observed velocity field (it was done
for the first time by Warner et al. (1973) for M~33 and Pence
(1981) for NGC~253). The attempts to relate the non-circular
velocities found in this way to the observed spiral structure were
unsuccessful, and this was treated by the authors as a failure to
find non-circular motion of wave nature in these galaxies.

On the other hand, it was known from earlier papers that there
exists a correlation between the velocity perturbations and the
pattern of spiral structure in some galaxies (see for example
Rubin et al. (1980) for optical rotation curves of galaxies or
Rots \&\ Shane (1975) for a HI velocity field of M~81).  If that
is the case, can the negative result obtained for M33 and NGC~253
be accounted for by some peculiarities in these spiral galaxies?
Indeed, although M~33 has two main spiral arms, it is really a
multi-armed galaxy (Sandage \&\ Humphreus 1980). In turn, NGC~253
is a starburst galaxy with a rather high disk inclination,
inappropriate for tracing spiral arms. However, later works by
Sakhibov and Smirnov (1987, 1989, 1990) made it evident that the
main problem of these early approaches is that independent
determination of the rotation curve and of the residual velocities
related to a density wave is not possible even for an ``ideal''
two-armed grand design galaxy. This conclusion was based on the
analysis of the line-of-sight velocity fields of some galaxies by
applying a Fourier-series expansion of velocity along the
galactocentric azimuthal angle. The authors mentioned above showed
that the even component of the first Fourier harmonic includes
contributions from both the rotational velocity and the motion
related to the density wave in the disk plane. Such interference
makes it necessary to work out a special method to separate the
circular and non-circular motion consistently. This method  was
proposed and described in detail in the papers by Lyakhovich et
al. (1997), Fridman et al. (1997, 2001). These authors also showed
that, in general, ignoring the influence of systematic
(density-wave related) non-circular motion onto the line-of-sight
velocities may prompt systematic errors in estimations of
orientation parameters and lead to a wrong conclusion about their
radial variations, even in the case of their constancy.

A similar problem appears when the photometric data is used to
determine $i$ or $PA_0$ of a disk.  In this case the isophote
distortion due to spiral arms may also mimic the radial variations
of orientation parameters. To avoid this difficulty the outer disk
isophotes are traditionally used where the contrast of spirals is
usually low and may be neglected. But in some cases these regions
may not be covered by the field of view, or are located too far
from the galactic center so that their orientation differs from
that of the inner disk (say, due to a disk warp). In this
situation the correct estimation of orientation parameters needs
a complex self-consistent analysis of brightness and/or of
line-of-sight velocity distributions, taking into account both
non-circular motion and brightness asymmetries, but this way is
rather cumbersome for practical use. A simplified method should
be chosen which may give an acceptable accuracy of orientation
parameters. As it will be shown below, this problem is not
trivial, and the use of different methods may lead to diverging
results.  Note that the accuracy of estimation of inclination and
position angle of a galaxy influences not only the rotation curve
obtained from observations. It may be a key factor which
determines the possibility and the reliability of the evaluation
of residual velocity field and of some refined parameters which,
for example, may be used to distinguish chaotic and regular
trajectories of gaseous clouds (Fridman, Khoruzhii \&\
Polyachenko 2002).

Section 2 contains descriptions of our observations and data
reduction procedures. In Section 3 we discuss various approaches
to the determination of orientation parameters from the kinematic
and photometric data and compare two independent methods, the
combined use of which enables us to solve the problem with good
accuracy. The parameters we found for 15 galaxies are used to
obtain their velocity curves in a model of pure circular motion
of gas. The discussion of individual galaxies is given in Section
4. Different types of non-circular gas motions are discussed in
Section 5. The main results of this work are summarized in
Section 6.

The detailed analysis of the systematic motions related to the
density waves, based on the method described above, will be given
in subsequent papers.

\section{Observations and data reduction}

The velocity fields of the chosen galaxies obtained in the
H$\alpha$ line cover significant parts of their discs, which
enables us to use them to compare the estimates of the
orientation parameters from both kinematic and photometric data
and to obtain the rotation curves of the galaxies.

The observations were made in 1995--2000 with the Scanning
Fabry-Perot Interferometer (IFP) attached to the reducer at the
prime focus of the 6m telescope of the SAO RAS; the equivalent
focus distance of the system is F/2.4. The description of the
interferometer can be found in Dodonov et al.  (1995).

We limit the required spectral range by using a narrow
interference filter with the $FWHM=10-20$~\AA~ and a
transmittance peak near a redshifted H$\alpha$ line of a galaxy.
When observing some galaxies (IC~1525, NGC~1084, etc.), the
emission lines [NII] $\lambda$ 6548, 6583 from the neighbouring
interference orders weakened by the wings of the filters also get
into the passbands. Fortunately, when using any of our
interferometers, the interference orders were overlaid in such a
way that the gap between H$\alpha$ and [NII] was not less than
half of the full spectral range, so the presence of [NII] in the
spectra did not prevent exact measurement of H$\alpha$, excluding
the case of the central part of NGC~5371 which is a LINER
possessing rather broad emission lines.

During various observational runs we used two IFP  by the
Queensgate company which are operating in 235th and 501st
interference orders (at a wavelength of 6562.78~\AA) with the
finesse of 8-12.  The parameters of the interferometers are given
in Table~\ref{ifppar}. The number of spectral channels (or steps
of scanning) for various objects was 24 or 32.  Also, various
types of panoramic detectors were used for different observational
runs. In 1995-1996 we used a 2D photon counter (IPCS) with a
format of $512\times 512$ pixels, in 1997 -- a CCD with a format
of $1040\times 1160$ pixels, from 1998 to the present -- the CCD
TK1024 with a format of $1024\times 1024$ pixels. In all cases
the observations were made with binning of $2\times 2$ pixels to
increase the S/N ratio and to economize the CCD readout time. The
data are presented as ``data cubes'' consisting of 24 or 32
object images obtained with varying gaps between the
interferometer plates.

For the different types of detectors we used different regimes of
scanning.  When observing with the IPCS, we accumulated an image
in every channel during 15--20 sec, and then the channel was
switched forward.  After passing through all the channels, the
scanning cycle was repeated.  This approach allowed us to overcome
variations of atmospheric transparency and of seeing quality
during total exposure. When observing with a CCD, each channel was
exposed only once, but to properly take into account variations
of atmospheric transparency, the odd channels (1,3,5...) are
exposed first, and then the even channels (2,\,4,\,6,\,...).

To correct spectra for the phase shift, we observe an emission
line of the calibrating lamp. When observing with the IPCS, the
calibration cubes were exposed before and after each object
exposure; when observing with a CCD -- at the beginning and at
the end of the night. To correct the data for the modulation
introduced by the order-separating filter and for the flat field,
the telescope main mirror cup or the wall of the dome lit by a
continuum light source are observed in every channel.

\begin{table}[t]
\caption[]{IFP  parameters.} \label{ifppar}
\begin{tabular}{lll}
\hline
                             &\multicolumn{2}{c}{Fabry-Perot interferometers
}
\\
                             &   FP235          & FP501
\\
\hline An order (at \Ha line)          & 235              & 501
\\
Interfringe ($\Delta\lambda$)& 28\AA ($1270\km$)& 13\AA
($600\km$)\\
Resolution ($\delta\lambda$) &3\AA ($130\km$)   & 1.2\AA
($130\km$) \\
Spectral sampling:           &                  &
\\
(24-channel mode)            &1.20\AA($53\km$) & 0.55\AA ($25\km$)
\\
(32-channel mode)            &0.90\AA ($40\km$) & 0.40\AA
($20\km$) \\
\hline
\end{tabular}
\end{table}

Reduction of the data obtained with the IPCS was performed in a
standard way with the software ADHOC (Boulesteix 1993). It
included a construction of the phase map by using the cube of the
comparison spectrum, flat-fielding, phase-correcting the object
cube (wavelength scale calibration), and sky spectrum subtraction
based on analysis of detector areas free of the object emission
and of its ghost images. To make an optimal filtration of the
data and to increase the S/N ratio in the regions with weak
emission, we smoothed each spectrum in the spectral domain by a 1D
Gaussian with $FWHM$ $=$ 2-3 channels and in the space domain by
a 2D Gaussian with $FWHM$ close to the seeing estimated from the
measurements of nearby stars.

The primary reduction of the data obtained with CCDs is performed
with the software developed by us and described in Moiseev
(\cite{mois02},\cite{mois02a}).  It includes bias subtraction,
flat-fielding, cosmic hit cleaning, and mutual shifts of frames to
overlay reference stars. The same stars are used to correct the
datacube, if necessary, for the variations of atmospheric
transparency and of seeing quality. Positions of the comparison
spectrum line are determined by fitting its profile by a Gaussian
or by a Lorenz profile; this modification has allowed us to
achieve somewhat higher accuracy in the phase map construction
with respect to the previous results obtained with ADHOC. After
night-sky spectrum subtraction and correction for the phase shift,
further reduction is performed by using the software ADHOC and a
new version of this software for WINDOWS, ADHOCw
(http://www-obs.cnrs-mrs.fr/ADHOC/adhoc.html).

\begin{table*}[t]
\caption[]{Log of IFP-observations.} \label{obslog}
\begin{tabular}{lrlllrlrrr}
\hline Name&   & Type(RC3)& Date  &Detector& IFP & px size  &
$T_{exp},
s$ & Seeing & Beam  \\
\hline
IC&1525 &SBb       &04 Dec 99& CCD   &FP235 & 0\farcs68&
$24\times180$& 2\farcs5 & 3\farcs2 \\
NGC& 23 &SB(s)a    &29 Aug 98& CCD   &FP235 & 0\farcs68&
$24\times180$& 2\farcs0 &2\farcs8  \\
NGC& 615&SA(rs)b   &29 Aug 98& CCD   &FP235 & 0\farcs68&
$24\times180$& 2\farcs5 &3\farcs8  \\
NGC& 972&SA(rs)b   &21 Feb 98& CCD   &FP235 & 0\farcs68&
$24\times180$& 2\farcs6 &3\farcs5  \\
NGC&1084&SA(s)c    &26 Oct 95& IPCS  &FP501 & 0\farcs92&
$32\times225$& 2\farcs0 &3\farcs5  \\
NGC&1134&S..?      &23 Feb 98& CCD   &FP235 & 0\farcs68&
$24\times150$& 2\farcs4 &2\farcs9  \\
NGC&2964&SAB(r)bc  &01 Dec 99& CCD   &FP235 & 0\farcs68&
$24\times180$& 2\farcs0 &2\farcs5  \\
NGC&3583&SB(s)b    &28 Feb 00& CCD   &FP235 & 0\farcs68&
$32\times180$& 2\farcs8 &3\farcs5  \\
NGC&3893&SAB(rs)c  &10 Dec 96& CCD   &FP501 & 0\farcs90&
$24\times180$& 2\farcs5 &3\farcs1  \\
NGC&4100&R'SA(rs)bc&21 Mar 96& IPCS  &FP235 & 0\farcs77&
$24\times220$& 2\farcs5 &3\farcs5  \\
NGC&4136&SAB(r)c   &21 Feb 98& CCD   &FP235 & 1\farcs36&
$24\times180$& 2\farcs5 &4\farcs0  \\
NGC&4414&SA(rs)c   &21 Feb 98& CCD   &FP235 & 0\farcs68&
$24\times120$& 2\farcs0 &3\farcs0  \\
NGC&4814&SA(s)b    &23 Feb 98& CCD   &FP235 & 0\farcs68&
$24\times120$& 2\farcs4 &3\farcs1  \\
NGC&5371&SAB(rs)bc &23 Mar 96& IPCS  &FP235 & 0\farcs77&
$24\times460$& 4\farcs0 &5\farcs5  \\
NGC&6643&SA(rs)c   &16 May 97& CCD   &FP235 & 0\farcs45&
$24\times360$& 1\farcs5 &2\farcs0  \\
\hline
\end{tabular}
\end{table*}

The journal of observations is given in Table~\ref{obslog}, where
for each galaxy we list its morphological type according to RC3,
the date of the observations, the detector and the interferometer
types which were used in each particular observation, pixel size
after binning, total exposure time in each channel, seeing
quality, and spatial resolution after all smoothing procedures
estimated by using completely reduced star images in the cubes.

\begin{figure*}
\centering
  \includegraphics[width=17cm]{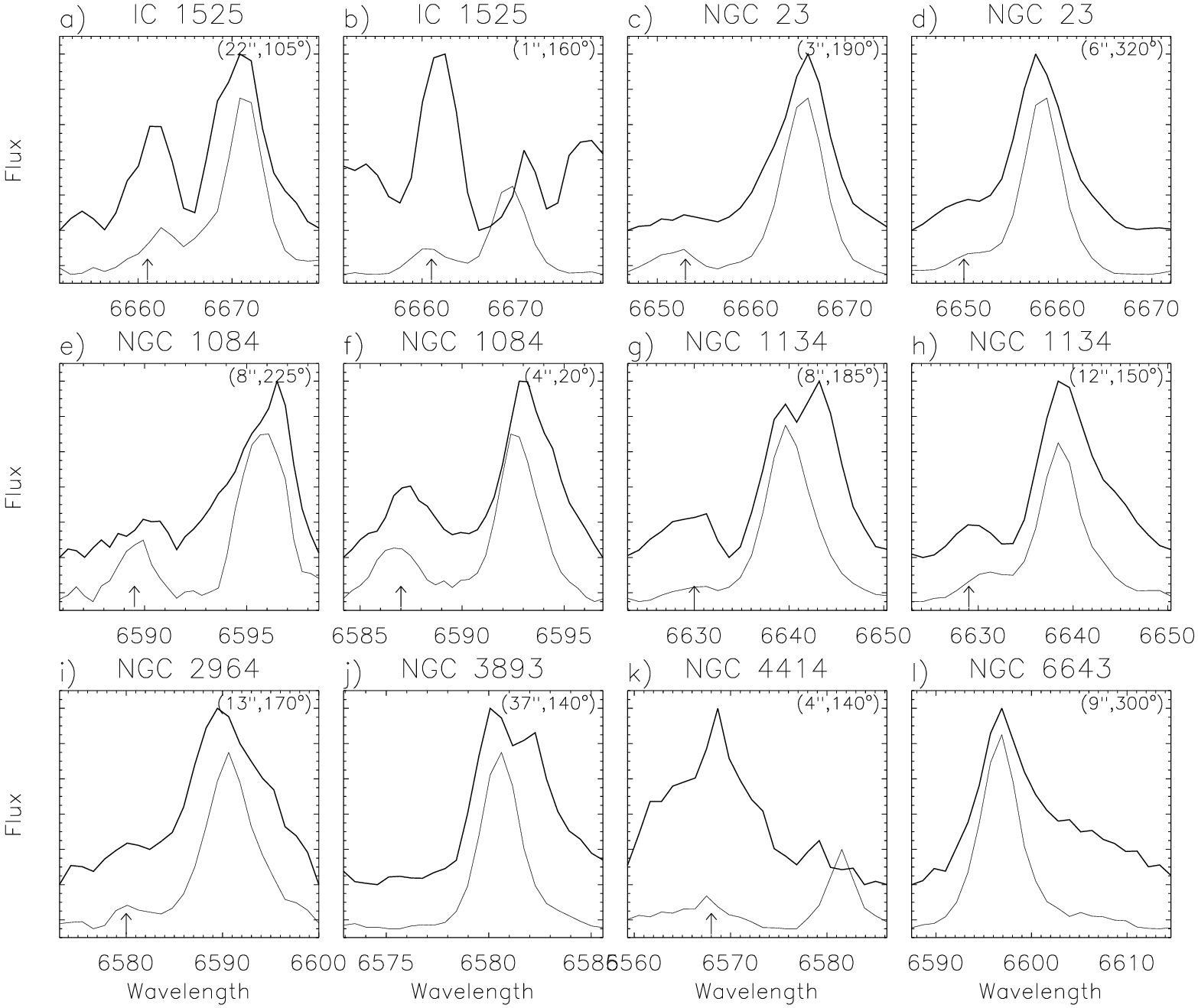}
\caption[]{ The examples
of the IFP-spectra with complex structure of the $H\alpha$
emission-line profiles (see Section~\ref{sec4} for the detailed
description of the individual galaxies). The thick line shows the
spectrum extracted for a region with the cylindrical coordinates
$(r,PA)$ given at the top-right corner of the each frame. These
profiles reveal different kinds of peculiarities, such as blue/red
wings (see c, d, e, f, h, i and l), double-humped structure (see
g, j), $H\alpha$-absorption distortion (see b and k), and the
[NII] emission line interference (see a, g, and h). The thin line
reproduces the spectrum of the nearest galactic region with
the``normal'' (symmetrical) shape of the emission line profile.
The arrows mark the position of the [NII] ($\lambda6548$\AA\, or
$\lambda6583$\AA) line, intruding from the neighboring
interference order. Note that the wavelength scale corresponds
only to the $H\alpha$ line velocities.} \label{fig_prof}
\end{figure*}

For each galaxy we have constructed a velocity field of the
ionized gas by measuring barycenters of the H$\alpha$ emission
line in each spectrum (for one-component line profiles), as well
as monochromatic images in H$\alpha$ and red continuum images in
the narrow passband near H$\alpha$. If the emission-line profile
is symmetric, the accuracy of the velocity measurements is
restricted only by the accuracy of phase map constructing and is
about $10-15 \km$ for the observations with the IPCS and about $5
\km$ for the CCD data. However in some cases (see below) the
emission lines are clearly double-humped or possess strongly
asymmetric profiles which is evidence for the presence of several
subsystems of ionized gas clouds on the line of sight. In these
cases we have performed a multi-component Gauss analysis of such
profiles to separate the kinematic components. The resulting
errors in the ``main component'' velocity measurements are about
$15-20\, \km$ for FP235 and $5-10\, \km$ for FP501 observations,
which is similar to the errors of barycenter-based velocities in
the case of a symmetrical line profile. To construct a velocity
field in the regions with multicomponent profiles we use only
those velocities which correspond to the ``main' component''. In
practically all the cases its value appears to be nearest to the
velocities found in the neighbouring regions with a one-component
shape of an emission-line profile. With the exception of a
complex situation in the central region of NGC~1084, such an
approach has not created any velocity discontinuity.  The
examples of the emission line spectra in the regions with the
``abnormal'' profiles are plotted in Fig.~\ref{fig_prof}.
Double-peaked profiles may be found in the circumnuclear regions
(NGC~23, NGC~1084, NGC~1134, NGC~2964, NGC~6643), as well as in
the outer disks (IC~1525, NGC~1084, NGC~3893).

The most complex case is represented by the circumnuclear regions
of galaxies with broad stellar H$\alpha$ absorption lines in the
spectra. Unfortunately, our spectral range is too narrow, and the
IFP order overlapping often prevents a proper correction for this
absorption feature. We met such situations in some regions in the
centers of IC~1525, NGC~4136, NGC~4414, and NGC~5371. We have
tried to use a double-Gaussian (emission + absorption) model to
calculate the velocities in these regions. However the errors of
the velocity measurements were too large ($\sim 50-70 \km $).
Therefore we have excluded these regions from the resulting
velocity fields (see the descriptions of the individual galaxies
in Section~\ref{sec4}).

\section{Results}
\label{sec3}

\subsection{``Kinematic'' orientation parameters of galactic disks}
\label{sec3.1}

After gas velocity fields of galactic disks became available
(first from HI radio line observations, later from optical line
measurements), several ways were proposed to solve the problem of
restoring rotation curves of galaxies from the obtained data. In
most cases the basic assumption is to consider deviations of the
observed velocities from those expected in a pure circular
rotation picture as random ones, which have a``noisy'' character:
\[
V^{\rm obs} ~=~ V^{\rm mod} ~+~ \Psi \, ,
\]
\begin{equation}
\label{eq:purerot} V^{\rm mod}(r,\, \varphi ) ~\equiv~ V_{\rm
sys} ~+~ V_{\rm rot}(r) \cos \varphi \sin i \, ,
\end{equation}
where $\Psi$ is a random deviation of the observed velocity from
that predicted in the model; $V_{\rm sys}$ is the systemic
line-of-sight velocity of the galaxy; $V_{\rm rot}$ is the model
circular velocity of the gas in the plane of the galaxy which
depends on the galactocentric radius $r$; $\varphi$ is the
galactocentric azimuthal angle measured with respect to the major
axis; $i$ is the inclination (the angle between the galactic
plane and the sky plane).

But even this very simplistic approach leaves the problem of
determining the rotation curve to be rather complicated. In
general, there are five parameters which should be found or taken
from elsewhere in order to extract the rotation velocities from
the line-of-sight velocity field.  They are: the systemic
velocity of a galaxy $V_{\rm sys}$, the position of the center of
mass $x_0, y_0$, the position angle of the line of nodes $PA_0$
and the disk inclination angle to the line of sight $i$.

In early papers (Warner et al. 1973, Pence 1981) these parameters
were assumed to be constant over the entire disk. In this case
they could be found either from the analysis of photometric data
(with the exception of $V_{\rm sys}$) (Pence 1981), or by
minimizing a sum of squared differences between the measured and
model-predicted velocities for a pure rotation model (Warner et
al. 1973, Pence 1981). In this case the rotation velocity at a
given galactocentric radius was calculated as a weighted mean of
all points lying in the plane of the disk within a ring of radius
$r$ and of width d$r$. Even in these early works it was revealed
that different methods give discrepant results, which was
ascribed to the influence of non-circular motion (Pence 1981).
Note that a similar effect could also be caused by non-planar
motion. As it was mentioned by van Moorsel and Wells (1985), the
Warner et al (1973) method is of a trial-and-error nature, which
makes it difficult to determine (cite)  ``what errors are to be
assigned to the results''. These authors proposed a modification
of the ring method assuming a certain functional approximation of
the rotation curve under the best-fitting orientation parameters;
after that they solved the system of many equations  for the
total number of unknowns  by a least-squares algorithm (without
using any weighting based on the position inside of a galaxy).
Unlike van Moorsel \&\ Wells, below we do not use any
parameterization of the rotation curve, applying the least-square
algorithm to the model of a thin circularly rotating disk where
the velocity of rotation of a given ring is considered as the
unknown.

The parameters $PA_0$ and $i$ estimated in the early works  were
found in some cases to be inapplicable to the outermost parts of
galaxies, where a distortion of isovelocities is often seen (see
for example Warner et al. 1973 for M~33, or Rogstad et al. 1979
for NGC~300). To improve the procedure without rejecting a
circular motion approximation, it was proposed (Begeman 1989) to
consider the values of $PA_0$ and $i$ (or in general, all the
parameters involved) as functions of $r$. This approach is called
``a tilted-rings model''. In this model it is assumed that the
disk of a galaxy may be described by a set of concentric rings
with fixed values of $V_{\rm sys}$, $x_0$, $y_0$, $i$, and $PA_0$
within each of them. In other words, each ring is allowed to have
arbitrary circular velocity and orientation parameters. The
values of $V_{\rm sys}$, $x_0$, and $y_0$ are often taken as the
same for all the rings. This approach is widely used to determine
rotation curves.\footnote{Note however that in some papers (see
Schommer et al. 1993, Amram et al. 1995,  Garrido et al. 2002,
2003) the authors  use constant values of orientation angles
$PA_0$ and $i$ for each galaxy as the best fit parameters, which
may be found in the frame of circular rotation model by the
least-square method or by iterations.}

Variations of the parameters are allowed not only at the periphery
of a disk, that may be physically advocated because of a possible
warp, but at all radii. It is worth mentioning that although the
warping of gaseous (and sometimes stellar) disks of galaxies is a
well-known fact, significant variations of inclination usually
take place in the outermost parts of galaxies and, as a rule,
beyond the regions covered by optical spectral measurements. It
seems that the parameter variations obtained by formal use of the
tilted-rings model is often a consequence of non-circular motions
which are not taken into account by this method.  If the
variations of either $i$ or $PA_0$ (or $x_0$, $y_0$) bear a
systematic character, it may be an indication that the presence
of a spiral density wave (or bar) should be taken into account
(Lyakhovich et al. 1997, Fridman et al. 1997).

As an illustration, Figs.6 \ref{vst_ok_1}\,a,~b show the radial
variations of $i$ and $PA_0$, formally determined by the
tilted-rings method for NGC~3893 -- one of the galaxies of our
sample with rather moderate inclination. The systemic velocity
and center position were assumed to be invariable.

\begin{figure*}
%
%
%
%
%
%
\caption[]{Radial variations of the inclination angle (left) and
position angle of the line-of-nodes (right) found by the
tilted-rings method applied to the observed velocity field of NGC
3893 (upper diagrams) and to the artificial galaxy (below), in
which the velocity field is a sum of pure rotation, non-circular
regular velocities expected in the simplified model of density
waves, and ``noisy'' velocity dispersion of about 10 km/s.  Thin
lines in the lower diagrams mark the values of parameters
obtained under the assumption of unique orientation parameters
for the whole velocity field. Circles are the estimates made for
the model in absence of density waves.  Thick lines are real
values of parameters accepted in the model.} \label{vst_ok_1}
\end{figure*}

In Figs. \ref{vst_ok_1}~c,~d analogous dependencies are shown for
an artificial (model) velocity field. This field was created as a
sum of pure rotation, non-circular regular velocities expected in
the simplified model of two-armed density waves, and a ``noisy''
non-correlated component imitating the errors of the velocity
measurements and the occasional small-scale velocity dispersion.
For definiteness, the amplitude of the rotational component was
taken to be $210 \km$, and of the regular non-circular velocities
-- $10-20 \km$, and the root mean square deviation of the
``noisy'' component was taken as $10 \km$. Circles mark estimates
of the parameters when regular non-circular velocities induced by
the density wave are excluded.  One can see that dependencies
obtained for the model which includes the density waves
reproduces the radial variations of parameters formally found for
the real galaxy rather well. It is also clear that when the
density wave is eliminated from the model velocity field, the
random motions may lead only to small parameter variations (by
1-2 degrees). It seems that the large variations of parameters in
the tilted-rings model (Figs. \ref{vst_ok_1}~a,~b) may be
naturally explained by the presence of a density wave.

The natural starting point for determining the orientation
parameters
 is an assumption that they remain constant over all
the optical part of a disk. But this in itself is not enough to
find the angles we seek in a correct way if regular velocity
perturbations in density waves take place.  The main problem is
that the latter introduce systematic, non-occasional deviations
from the pure circular motion, that do not allow the use of the
representation of the measured line-of-sight velocity in the form
(\ref{eq:purerot}). Due to the same reason, the orientation
parameters calculated for a given ring may be shifted in a
systematic, non-occasional way from their real values. It also
renders it useless to seek the most probable values by averaging
the data obtained for different rings, as it takes place often
(see for example Carignan et al. 1988, Begeman 1989, Beauvais \&\
Bothun 1999, 2000, 2001).

In order to not resort to the complex and laborious method of
constructing a self-consistent model of non-circular motion in
the disk, which being applied to an observational velocity field
may give the best fit parameters (see for example Fridman et al.
2001a, 2001b), one may use a simplified approach. The idea was
proposed by Fridman et al. (2001a), where a radial periodicity of
motion in the density wave leading to a periodical character of
the wave contribution into the Doppler velocity component is
suggested. Hence, if one is to seek the best values of the
orientation parameters, one can expect that among the methods
ignoring the density wave component of non-circular motion, the
best method uses the velocity field as a whole, without
estimating individual parameters for separate rings. In this case
the contributions of systematic deviations from circular rotation
in different parts of the disk are self-compensated, at least
partially.

Actually this approach (without discussing the role of
non-circular motion) was first used by Warner et al. (1973). Our
basic principles of the determination of rotation curves remain
the same as those used by Warner et al. (1973) aside from the
different approach to weighting the data.

A consistent and reliable way to estimate different components of
the Fourier spectrum of the line-of-sight velocity field (and
particularly the rotation velocity) was proposed and validated by
Lyakhovich et al.  (1997), Fridman et al. (1997, 2001a), Burlak
et al.  (2000). It is based on the application of a least-square
method to the expansion of the azimuthal distribution of the
observed gas velocities into the truncated Fourier series. The
rotational velocity is allowed to vary from one ring to another.
This is the only parameter which runs along the radius in our
model.

For a given ring of radius $r_l$ and width d$r$ a sum of squared
deviations of the measured velocities from those expected in the
model of pure circular rotation is:
\begin{equation}
\label{eq:leastsqr1} \chi ^2 (r_l) = \sum_{j}^{N_l} \left( V_j
(r_j, \, \varphi _j) - V_{\rm sys} - V_{\rm rot} (r_l) \cos
\varphi _j \sin i \right) ^2 \, .
\end{equation}

Here the summation is performed over all $N_l$ points of the ring
where the line-of-sight velocity was estimated, and the weighting
coefficients are assumed to be equal. \footnote{ We ignore a
weighting function in this work. The use of $\cos \varphi$ as a
weighting function is hardly justified here because the
parameters we are seeking are determined not from combining
individual velocity measurements, but rather from the functional
dependence of the line-of-sight velocities on the azimuthal angle,
the expression for which is supposed to be known. In this case
each point has a weight which does not depend on $\varphi$. See
also Begeman (1989) who wrote (cite): ``.. In principle there is
no objection to the use of equal weights, except when the
uncertainties in the determination of the radial velocities are
known to be different. (In general these uncertainties are not
known.) '' However he adds: ``The use of non-uniform [azimuthal]
weights [is] preferable...radial velocities in positions close to
the minor axis carry less information about the underlying
circular velocity than at positions close to the major axis. Also
by this reason one should give more weight to positions close to
the major axis''. Indeed the latter statement is correct only when
the circular rotation is restored from the individual local
velocity estimates, not from the comparison with the model
dependencies expected for a flat disc.} Analogous expressions may
be easily written for any choice of the weighting function. For
given values of the center position, systemic velocity,
inclination, and position angle of the line of nodes the minimum
of $\chi2 ( r_l)$ is reached under $\partial \chi2(r_l)/\partial
V_{\rm rot}(r_l)= 0$, that gives:
\begin{equation}
\label{eq:vrot1} V^{\rm rot} ( r_l) = \frac 1 {\sin i} \frac
{\sum_{j} \cos \varphi_j \left( V_j (r_j, \, \varphi_j) - V_{\rm
sys}\right) } { \sum_{j} \cos^2 \varphi_j} \, .
\end{equation}
For complete filling of the ring by the measurements, this
expression is reduced to a familiar one for the reverse Fourier
transformation.

Under condition (\ref{eq:vrot1}) the quadratic sum of deviations
in a ring (\ref{eq:leastsqr1}) takes the form:
\[
\chi^2(r_l) = \sum_{j} \left( V_j (r_j, \, \varphi_j) - V_{\rm
sys} \right)^2 ~-~
\]
\begin{equation}
\label{eq:leastsqr2} ~~~~~~~~~~~~~~~~\frac {\left[ \sum_{j} \cos
\varphi_j \left( V_j (r_j, \, \varphi_j) - V_{\rm sys} \right)
\right]^2 } { \sum_{j} \cos^2 \varphi_j} \, .
\end{equation}

For the whole disk we have:
\begin{equation}
\label{eq:leastsqr3} \chi^2(x_0,\, y_0,\, i,\, PA,\, V_{\rm sys})
= \sum_{l} \chi^2(r_l) \, .
\end{equation}

Some parameters contained in the brackets on the left side of
Eq.(\ref{eq:leastsqr3}), (say, the center position and/or
systemic velocity) may be considered as knowns. The others may be
found by minimizing the expression (\ref{eq:leastsqr3}). In
practice we estimated $V_{\rm sys}$ and the center position as the
best fit symmetry center of the observed velocity field by
comparing the velocities of points situated symmetrically with
respect to the center, and kept these values fixed.

If the rotation curve has a steep gradient in the central part of
a galaxy, it may be more convenient to use a slightly different
version of the expression for a quadratic sum of deviations. It
may be obtained in a similar way as written above, but starting
from a constant angular velocity for a given ring instead of a
constant linear velocity.  The resulting expression differs from
(\ref{eq:leastsqr2}) by the replacement of all values of $\cos
\varphi_j$ onto $x_j$ $=$ $r_j \, \cos \varphi_j$.

For the velocity field of the model galaxy which includes density
wave perturbations, the parameters $i$ and $PA_0$ found by the
method described above are shown by thin lines in
Figs.\ref{vst_ok_1}\,c,~d. They coincide with the parameters of
the model (thick lines) within a range of under one degree. It
enables us to conclude that the method gives correct estimates of
the orientation parameters, even in the presence of systematic
non-circular motions in a galaxy.

``Kinematic'' orientation parameters of our sample galaxies,
obtained by this method, are presented in the third column of
Table~\ref{orpar}. Their calculation was made by minimizing the
expression (\ref{eq:leastsqr3}) with fixed positions and systemic
velocities of the center. Confidence intervals given in
Table~\ref{orpar} were estimated from the formally determined
errors of the mean values in a model of circular rotation and
from the shape and width of the profile of $\chi2(i,\, PA_0)$
near its minimum. The obtained results are insensitive to the
chosen width of rings until it is smaller than the radial scale of
velocity variations in a galaxy. The $PA_0$ is traditionally
measured counter-clockwise from the North direction to the
receding part of the major kinematic axis of the galaxy.

Note that our method of interferometer data calibration may
introduce some small systematic errors in systemic velocity
estimation (however these errors do not influence the
measurements of relative velocity variations in the line-of-sight
velocity field).  A comparison of our $V_{\rm sys}$ (the second
column in Table~\ref{orpar}) with the NED database values shows
that the velocity difference is about $10-50 \km$ for different
objects -- that is of the same order as the scatter of estimates
obtained by different authors.

The fourth column of Table~\ref{orpar} contains the kinematially
determined orientation parameters obtained by a formal averaging
of values found for different rings (a classical tilted-rings
model). Because of the non-random variations of the orientation
angles of individual rings, it is incorrect to characterize the
confidence intervals of these estimates by the formal value of
the error in mean. For this reason, a root mean square deviation
from the mean value was taken as a confidence interval.  As
expected, this method of averaging gives estimates that are not
as reliable as those in the method we have preferred above.

\begin{table*}
\caption[]{Orientation parameters of galaxies determined by the
different methods.} \label{orpar} \fontsize{7}{9}\selectfont
\begin{tabular}{|c|r|r@{\hspace{8pt}}r@{\hspace{8pt}}c|r@{\hspace{
6pt}}r|c@{\hspace{6pt}}c|r@{\hspace{6pt}}r|c@{\hspace{4pt}}r|}
\hline
     &       & \multicolumn{3}{|c|}{Kinematics}  &
\multicolumn{2}{|c|}{Ring average}  &
\multicolumn{2}{|c|}{Photometry}&
\multicolumn{2}{|c|}{Leda} & \multicolumn{2}{|c|}{RC3} \\
\cline{3-13} Name   &  Vsys & $i,^\circ$ & $PA,^\circ$ & range,
$''$ & $i,^\circ$ & $PA,^\circ$ &  $i,^\circ$ & $PA,^\circ$ &
$i,^\circ$
& $PA,^\circ$ & $i,^\circ$ & $PA,^\circ$ \\
\hline IC 1525 & $5008\pm2$ & $43\pm5$ & $230\pm2$ & $r<90  $ &
$44\pm21$
& $230\pm4$ & $43\pm3$ & $210\pm5$ & $48$ & $200$ & $44$ & $200$\\
\hline
NGC 23  & $4553\pm3$ & $45\pm4$ & $174\pm2$ & $r<37  $ &
$48\pm18$
& $173\pm9$ & $45\pm3$ & $170\pm5$ & $43$ & $188$ & $50$ & $188$\\
\hline NGC 615 & $1875\pm3$ & $71\pm3$ & $156\pm2$ & $8<r<85$ &
$66\pm18$ & $153\pm15$& \raisebox{-\height}{$65\pm3$} &
$159\pm1,\,r<15''$ &
&       &        &     \\
        &            &          &           &          &
&           &          & $156\pm2,\,r>20''$ & $73$ & $205$ &
$66.5$ & $155$\\
\hline NGC 972 & $1512\pm2$ & $62\pm4$ & $325\pm2$ & $r<40  $ &
$53\pm23$ & $320\pm9$ & \raisebox{-\height}{$64\pm1$} &
\raisebox{-
\height}{$332^{+3}_{-1}$} & $65$ & $332$ & $59$ & $332$\\
        &            &          & $332\pm4$ & $40<r<45$&
&           &          &           &      &       &      &     \\
\hline NGC 1084& $1438\pm3$ & $56\pm3$ & $216\pm2$ & $r<60  $ &
$54\pm15$
& $215\pm4$ & $56\pm2$ & $217\pm2$ & $46$ & $295$ & $56$ & $215$\\
\hline NGC 1134& $3650\pm3$ & $53\pm4$ & $328\pm2$ & $r<50  $ &
$54\pm18$ & $326\pm5$ & $42\pm3,\,r<40''$ &           & &       &
&       \\
        &            & $67\pm3$ & $330\pm3$ & $50<r<90$&
&           & $67\pm2,\,r>50''$ & $326\pm3$ & $76$ & $328$ & $70$
& $328$ \\
\hline NGC 2964& $1320\pm4$ & $47\pm5$ & $278\pm2$ & $r<38$ &
$47\pm24$ &
$276\pm5 $ & $42\pm2$ & $276\pm2$ & $58$ & $277$ & $57$ & $277$\\
\hline NGC 3583& $2068\pm3$ & $46\pm6$ & $121\pm3$ & $r<70$ &
$45\pm24$ &
$122\pm12$ & $49\pm2$ & $124\pm2$ & $55$ & $125$ & $50$ & $125$\\
\hline NGC 3893& $953\pm3 $ & $42\pm4$ & $345\pm2$ & $r<70$ &
$43\pm18$ &
$345\pm3 $ & $40\pm4$ & $346\pm3$ & $32$ & $345$ & $52$ & $345$\\
\hline NGC 4100& $1150\pm3$ & $60\pm5$ & $358\pm3$ &$r<12   $&
&           & \raisebox{-\height}{$73\pm1.5$}& $353\pm2,\,r<10''$
&      &       &      &       \\
        &            & $72\pm2$ & $350\pm2$ &$25<r<90$& $63\pm17$
& $356\pm10$&           & $346\pm2$          & $78$ & $347$ & $71$
& $347$ \\
\hline NGC 4136& $586\pm2 $ & $25\pm9$ & $252\pm4$ &$r<95   $&
$33\pm16$ & $252\pm6$ & $18^{+5}_{-10}$ & $252\pm8$ & $22$ &
$179$ & $0-29$
&  --  \\
\hline NGC 4414& $742\pm2 $ & $56\pm4$ & $155\pm2$ &$r<90  $ &
$50\pm14$ & $156\pm2$ & $56^{+3}_{-1} $ & $157\pm3$ & $54$ &
$155$ & $56$ &
$155$\\
\hline NGC 4814& $2520\pm4$ & $50\pm5$ & $297\pm3$ & $r<90 $ &
$45\pm15$ & $297\pm6$ & $51\pm3 $ & $300\pm3$ & $46$ & $315$ &
$42$ & $315$
\\
\hline NGC 5371& $2618\pm5$ & $47\pm8$ & $195\pm4$ & $r<110$ &
$49\pm21$
& $193\pm10$ & $45\pm5$ & $195\pm3$ & $48$  & $188$ & -- &  -- \\

\hline NGC 6643& $1487\pm3$ & $61\pm4$ & $40\pm3 $ & $r<70 $ &
$54\pm17$ & $39\pm6  $ & $58\pm2$ & $37\pm2 $ & $57.5$& $38 $ &
$60$ & $38$
\\
\hline
\end{tabular}

\end{table*}

\subsection{``Photometric'' orientation parameters of galactic
disks}

\label{sec3.2}

Photometric methods are widely used because they are much simpler
and more understandable and do not demand analysis of the detailed
velocity fields. The methods are usually based on estimating
ellipticity and isophote orientation by fitting ellipses to the
outermost isophotes which can be reliably traced (they are
assumed to be undisturbed by disk warping or by internal
absorption). Another, more sophisticated way of getting the
results is to use azimuthal Fourier analysis of brightness
distribution of a galaxy in the plane of sky (Grosbol 1985) or in
the plane of the galactic disk (Iye et al. 1982).

As in the case of kinematic methods, the main problem of the
photometric approach is to take into account the deviations from
the axial symmetry caused by spiral arms or a bar. It may be
especially severe when a galaxy does not possess a radial zone
wide enough where the brightness distribution stays axisymmetric.
Ignoring this problem may lead to large systematic errors in
evaluating the orientation parameters.

In a formal sense, the second Fourier harmonic may appear as a
result of erroneous orientation parameters which are used for the
deprojection of an axially symmetrical disk. This harmonic also
appears if the parameters used are correct, but non-symmetrical
details such as spirals or a bar are present, so it is impossible
to separate these factors within the frame of Fourier analysis.

The only way to solve the problem is to use differences in the
phase and amplitude behavior of the second Fourier harmonic when
different factors are responsible for its appearance.
Particularly, the amplitude of the second harmonic caused by an
error in the estimation of the inclination grows as this error
increases. If the inclination is underestimated, the line of
maxima of the second harmonic would align with the major axis
because of this error, and in the opposite case it would coincide
with the minor axis. Unlike this behavior, the amplitude of the
second harmonic produced by a density wave should weakly
correlate with the orientation of the disk. Hence, to carry out a
Fourier analysis of the azimuthal brightness distribution under
initial assumption of zero inclination, and to draw the line of
maxima of the second harmonic, one can get the orientation of the
disk major axis (see also Grosbol, 1985, who used a similar
approach for the outer parts of real galaxies). The most clearly
delineated orientation of the line of maxima along the major axis
would be observed in the regions of a weak density wave (that is
for a poorly traced spiral structure or a bar), and even if the
width of each such region is rather small, coincident directions
of the lines of the harmonic maxima over all these regions enable
us to estimate an orientation of the major axis $PA_0$ with good
accuracy. It is essential that the method does not demand these
free-of-distortion regions to be preliminarily found in a galaxy:
they reveal themselves in the process of the second harmonic
phase calculations.

The inclination angle $i$ may be found (in the absence of the
non-axisymmetrical structure) as the angle where the phase of
maxima of the second harmonic of the deprojected image switches
from the major axis to the minor one. In the presence of a
density wave the interference of two factors (the error of $i$
and the influence of the spiral structure) causes the phase of
maxima of the second harmonic to be shifted into the position
between the major (or minor) axis and the observed brightness
maximum of a spiral arm in the given annulus. In the annuli where
the spiral arms have the lowest contrast (they may be revealed
during the procedure of $PA_0$ determination) the phase shift
should be the least. Let us note that when the probed inclination
passes through its correct value, the line of phase of the second
harmonic turns shortly in these regions because of the jump (at
$90^\circ$) of the position of the maximum contribution of the
error of $i$ in this harmonic. It allows us to estimate $i$ as an
angle where the phase of the second harmonic best fits the
observed spiral pattern (including the regions of low brightness
of the arms).

The photometric method described above has been successfully
applied to a sample of double-barred galaxies by Moiseev et al.
(2004); the reader may find some details of practical use of this
method in the cited paper. The results of ``photometric''
estimation of the orientation parameters for the present sample
galaxies are given in the fifth column of Table~\ref{orpar}. The
confidence intervals characterize an uncertainty caused by the
non-vanishing amplitude of the density wave. For the imaging data
we used (red) continuum brightness distributions obtained during
our observations with IFP. For three galaxies the published I-band
images obtained at the 1m ``Zeiss-1000'' telescope of the SAO RAS
were used. These are NGC~615 (Sil'chenko et al. \cite{oks}),
NGC~1084 (Moiseev \cite{mois}), and NGC~1134 (Bizyaev et al.
\cite{biz}). The R-band image of NGC~4136 obtained at the 1.1m
telescope of the Lowell Observatory was taken from the Digital
Atlas of Nearby Galaxies (Frei et al. \cite{frei}). For some
galaxies raw photometric data from the ING Archive of the UK
Astronomy Data Centre were also retrieved. These are the
observations of NGC~23 in R-band and NGC~6643 in I-band at the 1m
Jacobus Kapteyn Telescope and the observations of NGC~4100 and
NGC~6643 in R-band with the 2.5m Isaac Newton Telescope. The
K-image of NGC~972 was obtained with the UKIRT telescope (Hawaii)
and is kindly provided by Dr. S.D.Ryder.

\subsection{A comparison of ``kinematic'' and ``photometric'' estimates
of orientation parameters} \label{sec3.3}

It is convenient to compare the estimates of $i$ and $PA_0$
presented in Table~\ref{orpar} as follows: 1) by comparing our
``kinematic'' parameters obtained for the whole velocity field
with those taken as the average of the parameters found for
individual rings in a tilted-rings model; 2) by comparing  our
``photometric'' parameters with the inclination and position
angles contained in the LEDA and RC3 Catalogue, and 3) by
comparing the ``kinematic'' estimates with the ``photometric''
ones.

\subsubsection{The whole field versus ring average method.}

The confident intervals for the parameters $i$ and $PA_0$ obtained
as the means over individual rings are much larger than those
from our kinematic estimations. Although our whole-field
estimates formally fall into the wide confidence intervals of the
``tilted-rings model'' results, so that both could be considered
as ``consistent''\footnote{Formally speaking, two values $a\pm
\Delta_1$ and $b \pm \Delta_2$ are `consistent' if $|a-b|$ $\le$
$\sqrt{ \Delta_12 + \Delta_22}$.}, the accuracy of the parameters
obtained for the whole velocity field is significantly higher
than that using the average of parameters found for individual
rings in a tilted-rings model\footnote{We note above that it is
not correct to use the statistical error of the means as the
measure of the confidence interval when averaging over individual
rings. However, if one does it formally, the difference between
the angles estimated for the whole field and by averaging over
rings would be greater than the confidence interval.}.

\subsubsection{Photometric method taking into account spiral
structure versus outer-isophote methods.}

An agreement between our estimates of $i$ and $PA_0$ obtained by
the photometric method taking into account the regular spiral
structure with those listed in the LEDA database and RC3
catalogue appears to be rather moderate. Note however that the
accuracy of the LEDA and RC3 data is certainly lower than ours as
they are based on small-scale Palomar images of galaxies.
Concerning LEDA, one of the parameters $i$ and $PA_0$ are
``consistent'' (within $\sqrt{2} \Delta_1$) for about a half of
the objects, and only for 3 galaxies (NGC~972, 4414, and 6643) the
``consistency'' exists for both orientation parameters. A
comparison with the RC3 catalogue gives better results:
inclination estimates are ``consistent'' for 8 of 14 and $PA_0$
-- for 10 of 13 common galaxies. Both parameters are `consistent'
simultaneously for 6 galaxies.

Nine galaxies of our sample are also included in the list of
objects for which Grosbol (1985) estimated the orientation
parameters by analyzing the second Fourier harmonic of the
azimuthal brightness variations on small-scale Palomar images at
several radial bins (IC 1525, NGC 23, 2964, 3583, 3893, 4136,
4414, 4814, 5371). The root mean square difference between his
values and ours obtained by the photometric method is $6^\circ$
for $i$ and $11^\circ$ for $PA_0$. The root mean square
differences between Grosbol's and LEDA data are slightly higher,
$8^\circ$ for $i$ and $12^\circ$ for $PA_0$ (even if to exclude
NGC~4136 with strongly discrepant estimates of $PA_0$).

\subsubsection{Kinematic versus photometric parameters.}

A comparison of the estimates of $PA_0$ we obtained by the
kinematic and photometric methods described in the previous
subsections reveals discordant results in 3 of 15 galaxies
(IC~1525, NGC~972, and NGC~4100). For the first galaxy the
disagreement is most likely due to the presence of a rather
strong bar. For NGC~972 the discordant estimates may also be
related to the bar. Although this galaxy is commonly classified
as a non-barred one, the presence of the inner bar is found in
the infrared (see the next section). In NGC~ 4100 the difference
of $PA_0$ found by two methods is rather low (if to exclude the
inner part, the kinematic and photometric $PA_0$s are $350\pm 2$
and $346 \pm 2$ degrees respectively) and may be coincidental.

The non-circular motion caused by spiral density waves can
distort our kinematic estimates of $PA_0$, but these distortions
are different in different parts of the disk and as a result they
partly compensate each other when the whole velocity field is
taken for the $PA_0$ determination. The presence of a bar may
produce stronger non-circular motion with the preferred
orientation that are hard to compensate by averaging. It may
explain the differences between the orientation parameter
estimates obtained in some barred galaxies by the two methods.
Figs. \ref{vst_ok_3} and \ref{vst_ok_4} demonstrate the shapes of
the lines of maxima of the second Fourier harmonics of the surface
brightness for the barred galaxies IC~1525 and NGC~972 calculated
using different estimates of $PA_0$. As one may see, the shape of
the lines calculated with the $PA_0$ kinematically estimated
(left) disagrees with the observed spiral arms. Instead, these
lines tend to oscillate around some preferred direction which is
close to the orientation of the major axis according to the
photometric estimate of $PA_0$. The shapes of the lines
calculated with the latter $PA_0$ values (right) are in
accordance with the spiral arms traced by eye. All these
considerations allow us to suggest that the true orientation
parameters of these two barred galaxies should be closer to those
derived from the photometric data.

\begin{figure*}

%
%
%
%
\caption[]{A position of maxima of the second Fourier harmonic of
the surface brightness calculated for different choice of $PA_0$
superimposed onto the image of IC~1525. Left (a, c): $PA_0$ is
chosen according to the kinematic data analysis. Right (b, d):
this parameter was taken from the photometry analysis (see the
text). Images are in the red continuum (above) and in H$\alpha$
line (below).} \label{vst_ok_3}
\end{figure*}

\begin{figure*}
%
\caption[]{A position of maxima of the second Fourier harmonic of
the surface brightness calculated for a different choice of $PA_0$
superimposed onto the K-band image of NGC~972. Left (a): $PA_0$
is chosen according to the kinematic data analysis. Right (b):
this parameter was taken from the photometry analysis.}
\label{vst_ok_4}
\end{figure*}

Concerning the inclination $i$, its values found by us from
kinematic and photometric data are ``consistent'' in 13 galaxies
of 15. Those two galaxies (NGC~615 and NGC~1134) where the
agreement is absent demonstrate rather complex asymmetric
structure.

Indeed, we would advise use of both methods together when trying
to estimate galaxy orientation parameters: a convincing reason is
that the non-axisymmetry of a galactic disk -- a bar or a strong
density wave -- affects the parameter estimations by photometric
and kinematic methods in a different way.

  \subsection{Rotation curves in a model of pure circular gas
motion} \label{sec3.4}

Fig.\ref{ic1525} shows the results of data processing for the
galaxy IC~1525. Similar illustrations for other galaxies of our
sample are presented in the electronic version of the article.
\begin{figure*}[h!]
\centering
\caption[]{IFP data for
galaxy IC~1525. Top -- Image in continuum near H$\alpha$ (left)
and monochromatic image in H$\alpha$ (right), a square-root scale
is used. Bottom left -- velocity field, the thickest black
contour corresponds to the systemic velocity, other contours are
$\pm50, ~\pm100, ~\pm150$ ... $\km$. A cross marks the dynamical
centre. Bottom right -- rotation curve $(V_{\rm rot})$, kinematic
position angles $(PA)$, and the inclination $(i)$ (both are
obtained in a tilted-rings model). Dashed lines show the accepted
kinematic parameters $PA$ and $i$. The error-bars on the $V_{\rm
rot}(r)$ curve correspond to the mean non-circular velocities
(see the text for details).} \label{ic1525}
\end{figure*}
 For each galaxy the first three plots show the images in the red
continuum and H$\alpha$ (gray-scaled in proportion to the square
root of the intensity), and line-of-sight velocity field of
ionized gas with the isovelocity lines superimposed with a step of
$50\km$ (the thick line corresponds to the systemic velocity
$V_{\rm sys}$).
 The coordinates of the center of a galaxy and its systemic
velocity were found from kinematic data as the point of the best
mirror symmetry of the velocity fields for all pairs of pixels
taken on opposite sides with respect to the center. A small
variation of the center coordinates has little effect on the
orientation parameter estimates because they are influenced by
different Fourier harmonics of the velocity field (see Lyakhovich
et al. 1997). Additionally, we have determined the positions of
the center from both photometric and kinematic data independently
for several galaxies to be sure that they coincide within 1-2
pixels and it is always the case.

In the last plot the rotation curve is shown (the upper curve).
It was calculated from the observed velocity field (see
Eq.~\ref{eq:vrot1}) as the best fit curve of circular rotation for
the ``kinematic'' orientation angles $i$ and $PA_0$, which were
assumed to be constant along the radius. The lower curves are the
variations of $i(r)$ and $PA_0(r)$ obtained for the individual
rings in a tilted-rings model.

Since each elliptical ring contains from several tens to several
hundreds of measured points, formal errors of the mean $V_{\rm
rot}(r)$ in the most cases are lower than $1 - 5\km$. However, as
it was pointed out by Lyakhovich et al. (\cite{lyak1997}) and
Fridman et al.  (\cite{frid1997}), the rotation curve obtained in
the frame of the model of pure circular motion may have a
systematic error of the order of non-circular velocities in a
density wave. For this reason, with the error bars we mark the
mean residual velocity in the individual rings
($\sqrt{\chi^2(r_l)/N_l}$, see Eq.~\ref{eq:leastsqr2}). This way
is evidently more correct than using formally determined errors
of $V_{\rm rot}(r)$ because it demonstrates a real confidence in
these estimates. Genuine circular rotation curves, which may be
found by methods taking into account the regular non-circular
motion, will be within the error bars of the curves presented
here. The large error bars seen at some rotation curves give
evidence for significant deviations from the circular motion
model rather than for the low accuracy of measurements.

\section{Results for individual objects} \label{sec4}

Figures illustrating the results obtained for all galaxies of our
sample but the first one, IC~1525, are available in the
electronic version of the paper.

\subsection{IC 1525}

 Below we will use $PA$ and $R$ as polar
coordinates in the sky plane (where $PA$ is the position angle
measured from the North counterclockwise and $R$ is the distance
from the galactic center).

The continuum image (Fig. \ref{ic1525}\,a) reveals the presence
of a prominent bar aligned at $PA=48\degr$. The bar ends are
encircled by the ring of HII regions which has a radius of
15\arcsec -- 17\arcsec~ (Fig.~ \ref{ic1525}\,b). In the central
part of the bar, $R< 5\arcsec -7\arcsec$, due to the
superposition of strong absorption lines of bulge stars, the
H$\alpha$ emission line sinks in absorption, so the velocity
field (Fig.~ \ref{ic1525}\,c) has a hole in the center.

During the observation of this galaxy, the working spectral range
of IFP contained not only H$\alpha$, but also the emission line
[NII]$\lambda$6583 from the neighboring interference order, the
most prominent line in the circumnuclear region (see
Fig.~\ref{fig_prof}\,b). We have tried to estimate the
line-of-sight velocity of the ionized gas in the central region
lacking the H$\alpha$ emission by using the [NII] line, but the
velocity field obtained in such a way deviates strongly from the
simple extrapolation inward of the mean rotation field obtained
from the H$\alpha$ measurements further from the center. We think
that the anomalous velocities obtained from the [NII]
measurements in the region $r<5\arcsec$ may result from the
impossibility to properly take into account the effect of the
stellar absorption line H$\alpha$ disturbing the profile of [NII]
emission line because of the overlapping  of interference orders.
To treat the gas motion in the center of IC~1525 correctly, new
observations of emission lines free of stellar absorptions are
needed.

The residual velocity distribution, obtained by the subtraction
of the mean rotation field from the observed velocity field,
demonstrates several regions where a projection of non-circular
velocities onto the line of sight exceeds $50 \km$. The sizes of
these regions corrected for the beam smearing are $1\arcsec -
5\arcsec$~ (up to 1.6 kpc for the adopted distance of 67 Mpc). In
the cylindrical coordinate frame on the sky plane, at the locus
$R=28\arcsec$, $PA=190\degr$ the residual line-of-sight velocity
is $50 - 80 \km$; at the loci ($R=33\arcsec$, $PA=34\degr$),
($R=22\arcsec$, $PA=105\degr$), and ($R=35\arcsec$, $PA=75\degr$)
the residuals have the opposite sign:  $-(40 - 70) \km$. It is
interesting that the intensity ratio [NII]/H$\alpha$ is enhanced
in these regions by a factor of 1.5-2 with respect to the nearest
outskirts of these local velocity anomalies (see
Fig.~\ref{fig_prof}\,a). At the locus ($R=26\arcsec$,
$PA=254\degr$) the relative intensity of [NII] is quite normal,
but the width of the H$\alpha$ emission line profile is $FWHM=200
- 220 \km$, (after deconvolution with the instrumental contour),
and the residual line-of-sight velocities are $-(50 - 80) \km$.
All five sites of anomalous velocities are located beyond bright
HII regions and are not related spatially to the spiral arms.

A sharp steepening of the very outer part of the velocity curve in
this and some other galaxies discussed below is evidently an
artifact; the result of a small number of reliably measured
points (often located near the minor axis of a galaxy) and/or of
a bad agreement with a model of co-planar circular motion.

\subsection{NGC 23 (Mrk 545)}

The rotation curve of the galaxy (Fig.~\ref{ngc23}\,d) possesses a
local maximum first noted by Afanasiev et al. (1988a). We have
prolonged this rotation curve to the radius twice as large as that
reached by Afanasiev et al. (1988a). Since at $r>30\arcsec$ there
is only one bright ``spiral arm'' that contributes to the rotation
velocity measurements, the local minimum of the rotation curve at
$r\approx 40\arcsec$ is a possible artifact related to
non-circular motion within this arm. The inclination of
individual rings in a tilted-rings model vary strongly along the
radius that evidently reflects significant deviations from
circular motion of about $20-30\km$, nevertheless the inclination
$i$ and the position angle of line-of-nodes $PA_0$ we found for
this galaxy (see Table~\ref{orpar}) remain to be consistent
within their error with the earlier photometric and kinematic
estimates (Afanasiev et al. 1988a). The $PA_0$ of the kinematic
major axis orientation in the inner part of NGC~23 also coincides
within a few degrees with the estimate of Afanasiev et al.
(1991b) made with the integral field (multipupil) spectrograph
MPFS.

The problem of possible bar presence in NGC~23 remains unsolved.
According to RC3, the galaxy is classified as SB, and de Jong
(1996), by applying a 2D decomposition method to its brightness
map, finds a bar with a length of $29\arcsec - 32\arcsec$ aligned
along $PA=154\degr$. But the velocity field obtained by us does
not demonstrate any noticeable turn of the kinematic major axis
(Figs.~\ref{ngc23}\,c and \ref{ngc23}\,d), and there is also no
sign of any isovelocity twisting when passing from the outer to
the inner region of the galaxy (Afanasiev et al. 1991b). The
local small-scale twisting of the isovelocities at $r=5\arcsec -
8\arcsec$ (Fig.~\ref{ngc23}\,d) is rather related to a small
(mini- ) bar with a length of about 10\arcsec, which is implied by
the CCD- observations of Chapelon et al. (1999).

In the circumnuclear region ($r<5\arcsec$) the emission line
H$\alpha$ is broadened up to $FWHM=200 - 300 \km$, which was
already noted by De Robertis \&\ Shaw (1988) from their
high-resolution spectroscopy. Their Fig.~1 demonstrates two local
maxima of $FWHM$ located symmetrically with respect to the
nucleus at $r=6\arcsec - 8\arcsec$. Our IFP spectra of these
spots reveal strong either red or blue wings of the H$\alpha$
line profiles. The Gauss analysis of these profiles allows us to
derive ``non-circular'' velocity components shifted with respect
to the main component by $130 - 200 \km$ at ($R=6\arcsec$,
$PA=323\degr$) and by $-(170 - 200) \km$ at ($R=3\arcsec$,
$PA=190\degr$). The profiles of the emission lines in these
regions are shown in Figs.~\ref{fig_prof}\,c and \ref{fig_prof}\,
d. The estimated diameters of these regions are $3\arcsec -
4\arcsec$, being close to our spatial resolution element. We
should note that it is at the location of the ``blue'' component
where Afanasiev et al. (1991b) had found non-circular motion of
the ionized gas with velocities of $260-300 \km$, but the
asymmetry of the emission line profiles was missed by them
because of their low spectral resolution. The same authors have
argued that NGC~23 may be at the post-Seyfert stage. In this case
the anomalous gas motion may be related to the activity, or to a
starburst in the nucleus. But the emission line intensity ratios
in the nucleus of NGC~23 evidence rather in favor of HII-region
character of excitation and are not consistent with the
hypothesis of Sy2 or LINER (De Robertis \&\ Shaw 1988; Contini et
al. 1998). Since the sites of the anomalous velocities are
located at the ends of a nuclear minibar (see above), the
presence of the second velocity component may be related to
non-circular motion in the bar. But in this case one needs to
explain a very high degree of non-circular motion -- more than
$100 \km$ in projection along the line of sight.

\subsection{NGC 615}

In the circumnuclear region ($ r < 10\arcsec-15\arcsec$) the broad
absorption line H$\alpha$ is superposed on the emission line
[NII]$\lambda$6548 from the neighboring interference order that
prevents an exact determination of the line-of-sight velocities
even by a Gaussian analysis. We start our measurements beyond the
radius of the circumnuclear local rotation velocity maximum (at
$r\approx 5\arcsec$) detected by Afanasiev et al. (1988a). Beyond
$r=10\arcsec$ the relative intensity of the nitrogen emission
lines falls, so they do not affect the profile of H$\alpha$ any
more.

The most prominent feature of the monochromatic H$\alpha$ image
(Fig.~\ref{ngc615}\,b) is a pair of HII regions located
symmetrically with respect to the nucleus at the radius of
$20\arcsec - 23\arcsec$. They are slightly elongated along the
radius and look like shock fronts at the edges of a bar (though
the galaxy is classified as unbarred). At the same distance from
the center the gas velocity field is disturbed by non-circular
motion resulting in a turn of the kinematic major axis by more
than $20\degr$ at $r=15\arcsec - 30\arcsec$. We have found here
several spots with the line-of-sight velocities differing by $40
- 50 \km$ from the mean rotation field. These kinematic details
may be evidence for a triaxial potential within $r=25\arcsec
-30\arcsec$ in this galaxy. Indeed, the recent photometric
analysis of Sil'chenko et al. (2001) has revealed that NGC~615
possesses a separate inner compact disk of oval shape especially
notable at $r=15\arcsec - 30\arcsec$ where it introduces
non-monotonic changes of isophote orientation and ellipticity.

\subsection{NGC 972}

The near-H$\alpha$ continuum image (Fig.~ \ref{ngc972}\,a) lacks a
spiral structure, but on the H$\alpha$ image
(Fig.~\ref{ngc972}\,b) the HII regions concentrate toward a
pseudo-ring with the radius of about 30\arcsec. Several works
which analyze CCD-data on NGC~972 (Ravindranath \&\ Prabnu 1998,
Zasov \&\ Moiseev 1998) also contain remarks about the absence of
a noticeable spiral structure on the broad-band optical images
and about a large amount of dust within the ring of HII regions.
However, near-infrared images obtained through $JHK$-filters
reveal a prominent two-armed spiral pattern and a small bar with
the size of 10\arcsec\ elongated in $PA=175\degr - 180\degr$
(Mayya et al.  1998; Zasov \&\ Moiseev 1998). This peculiarity
makes the morphological classification of the galaxy very
ambiguous; e.g. Mayya et al. (1998) classify it as SABdpec while
according to NED (RC3) it is Sab.

The isovelocities of the ionized-gas velocity field within the
central 10\arcsec\ have a characteristic S-shape
(Fig.~\ref{ngc972}\,c), evidently induced by a presence of the bar
seen in NIR. A turn of the kinematic major axis in the center
exceeds $30\degr$ (Fig.~\ref{ngc972}\,d); the non-circular
(radial?) gas streams in the bar are so significant that a
cross-section of the velocity field in $PA\approx 30\degr$ gives
an impression of counterrotation. We should note here that the
presence of a minibar in the center of NGC~972 had been first
claimed by Zasov \&\ Sil'chenko (1996) from kinematic arguments.

The orientation parameters found by us for the outer part of the
disk (Fig.~\ref{ngc972}\,d) are consistent within the errors with
both photometric (Ravindranath \&\ Prabnu 1998, Zasov \&\ Moiseev
1998) and kinematic results from our earlier long-slit
observations (see Zasov \&\ Moiseev 1998, who used the data from
Afanasiev et al. 1991a). The rotation curve derived in this work
also agrees rather well with that obtained by Zasov \&\ Moiseev
(1998).

Ravindranath \&\ Prabnu (1998) noted a high rate of star
formation in the central part of NGC~972, comparable even to that
of the well-known starburst galaxies M~82 and NGC~253. We see
indirect consequences of this intense star formation when
analyzing our Fabry-Perot spectra:  many HII regions stand out by
their negative residual velocities of about $-(15 - 20) \km$,
which is probably related to expanding bubbles of ionized gas. A
negative sign of residuals may be accounted for by the presence
of internal
 absorption, dimming the radiation of receding parts of the bubbles.
In the nucleus, at $r<2\arcsec - 3\arcsec$, which is close to our
spatial resolution, the H$\alpha$ emission-line profile has
mostly a two-component structure; it can be related either to gas
outflow from the starburst nucleus or to non-circular gas motion
at the edges of the minibar (analogous to NGC~1084, see below).

\subsection{NGC 1084}

The emission line [NII]$\lambda$6583 from the neighboring
interference order is seen throughout the galaxy; it has allowed
us to study the kinematics of the ionized gas by using two
emission lines separately instead of a single H$\alpha$. From the
earlier observations by Afanasiev et al. (1988a), two regions
with strong velocity deviations from the mean rotation were
found. Particularly in the central region, $r<5\arcsec$, the
rotation velocities obtained by measuring the forbidden emission
lines [NII] and [SII] were systematically lower than those
obtained by measuring H$\alpha$. The authors had interpreted this
peculiarity as evidence for minibar
 presence in the center of NGC~1084. Besides that, non-circular
motions were found at the periphery of the galaxy, at $r=40\arcsec
- 50\arcsec$ to the north-east from the center, near an elongated
arc-like spur outgoing from the spiral arm near a giant
superassociation (Fig.~\ref{ngc1084}\,b).

Our IFP data confirms the existence of these kinematically
distinct regions. Near the spur, the profile of the H$\alpha$
emission line is mostly two-component: one component traces the
mean rotation and the other is shifted by $\pm 100 - \pm 150
\km$. The appearance of the non-circular component is
supplemented by a sharp increase of the [NII]/H$\alpha$ ratio by
a factor of $2 - 6$ with respect to the normal one in the
neighboring HII regions. The areas of non-circular motion have
typical sizes of 1-2 kpc and are located between the HII regions.
A possible interpretations of these non-circular motions of
ionized gas near the spur is discussed elsewhere (Moiseev 2000).

In the circumnuclear region, $r<5\arcsec$, the profiles of the
H$\alpha$ and [NII] emission lines also demonstrate a
two-component structure, with the velocity separation of the
components about $50\km$ ( Figs.~\ref{fig_prof}\,e and
\ref{fig_prof}\, f). The Gaussian analysis of the H$\alpha$
profile has allowed us to trace azimuthal variations of the
observed velocity gradients for each component. We have found
that the brighter component traces the circular rotation of the
ionized gas with the kinematic major axis close to the major axis
of the galaxy, whereas the fainter component of both lines
corresponds to the gas radial motion because its projected
velocity gradient is zero near the line of nodes.

In the distance range of 5\arcsec-20\arcsec~ on the opposite sides
of the nucleus, in the direction $ PA\approx 60\degr$, the
extended linear front is revealed. At both sides of this front
the character of asymmetry of the H$\alpha$ and [NII] profiles
changes abruptly: the red wings are replaced by blue ones. The
line-of-sight residual velocities of the main line components
change their sign across the front jumping by $80 - 100 \km$. The
width of the front is $3\arcsec - 4\arcsec$, so it remains
practically unresolved. Such a combination of features evidences
that we are dealing with supersonic gas flow in a strong shock
front.

Features similar to those observed in NGC~1084 (radial gas flows,
narrow shock fronts) are usually treated as a result of bar
influence; but neither our continuum image of NGC~1084
(Fig.~\ref{ngc1084}\,a), nor broad-band photometric images of the
galaxy (Zasov \&\ Moiseev 1999, Moiseev 2000) reveal any
signatures of a bar. We should also note that the turn of the
kinematic major axis in the radius range of the shock fronts
($5\arcsec - 20\arcsec$) is rather small (Fig.~\ref{ngc1084}\,d),
which implies the absence of a high contrast triaxial potential.

A whole complex of peculiarities observed in the center of NGC
1084 may be interpreted as a result of a weak triaxial potential
(a small triaxial bulge). Similar shock fronts at the edges of a
slightly triaxial bulge have been also observed in Sb galaxy
NGC~2841 by Afanasiev \&\ Sil'chenko (1999), but in that galaxy
the bulge is much more massive.

\subsection{NGC 1134}

Due to its complex inner structure (multiple dust filaments, etc.)
the galaxy is included in Arp's Atlas of Peculiar Galaxies (1966),
though it is not an interacting system. In the area $r<20\arcsec
- 30\arcsec$, HII regions trace a three-arm spiral
(Fig.~\ref{ngc1134}\,b), but in the outer part of the galaxy the
spiral pattern is two-armed. The three-arm spiral structure in the
center of NGC~1134 can also be seen when analyzing its direct
images by constructing maps of a color Q-parameter free of dust
influence (Bizyaev et al. 2001).

The mean orientation of the line of nodes of the disk
(Fig.~\ref{ngc1134}\,d) is consistent with the earlier photometric
and kinematic estimates (Afanasiev et al. 1991b), though our
estimate of the inclination is significantly lower than
$i=72\degr$ used by these authors. At small radii, $r<12\arcsec$,
the kinematic major axis (Fig.~\ref{ngc1134}\,d) turns abruptly
by more than $15\degr$, so that the isovelocities
(Fig.~\ref{ngc1134}\,c) demonstrate an S-like shape typical for a
bar potential. Above all, on both sides of the nucleus, at
$r=4\arcsec - 7\arcsec$ there are regions where H$\alpha$
emission-line profiles are two-component, with a separation
between the components of $150-200 \km$ ( see
Figs.~\ref{ngc1134}\, g and \ref{ngc1134}\, h. These regions of
the two-component H$\alpha$ emission line are elongated along the
direction of $PA\approx 20\degr - 30\degr$ and resemble the shock
edges in NGC~1084. Close to these shock fronts, at the minor axis
of the galaxy, one can note two spots where the secondary velocity
component disguises itself as an asymmetric wing of the H$\alpha$
emission-line profiles, and the velocities obtained for the peak
of profiles differ from the mean rotation velocities by $\pm (80
- 100) \km$. Since the anomalous velocities are observed near the
minor axis of the galaxy, they may be explained as the presence
of radial gas flows. We should mention that for the first time
the regions of the non-circular gas motion on both sides of the
nucleus in NGC~1134 were detected by Afanasiev et al. (1991b).

All the kinematic peculiarities of the central region of NGC~1134
mentioned above can be explained by the hypothesis of a bar
aligned with the minor axis of the galaxy.

The mean rotation curve calculated by us (Fig. ~\ref{ngc1134}\,d)
agrees with the one reported earlier by Afanasiev et al. (1991b):
the solid-body area up to $r=7\arcsec$, a low-contrast peak, and
after that a velocity plateau. At $r=65\arcsec - 75\arcsec$ and at
$r=90\arcsec - 110\arcsec$ there are two emission ``islands'' in
the disk, emerging by $70 - 90 \km$ from the extrapolated flat
velocity of rotation, which are responsible for the sharp jumps
in the calculated velocity curve at large distances from the
center. These anomalous details can be related either to a
gas-disk orientation change at large radii (a ``warp''), or to the
possibility that these HII regions do not belong to the global
disk plane, being dwarf satellites of NGC~1134. To clarify these
possibilities, the results of HI observations of the outer disk
kinematics are needed.

\subsection{NGC 2964}

The galaxy possesses a well-developed two-arm spiral structure;
HII regions at $r=30\arcsec - 35\arcsec$ are concentrated to a
pseudo-ring (Figs.~\ref{ngc2964}\,a and \ref{ngc2964}\,b). It has
a close companion, NGC~2968, at the distance of $5\farcm 8$, but
there are no evident signatures of interaction of galaxies.

At $r=1\arcsec - 5\arcsec$ the kinematic major axis
(Fig.~\ref{ngc2964}\,d) turns abruptly by more than $15\degr$,
which can be assumed to be the effect of a small bar (note that
there is a remark about the ``very faint bar'' in RC1). The
high-resolution image of the central $18\arcsec \times 18\arcsec$
region of NGC~2964 obtained by HST and published by Carollo et al.
(1997) reveals the presence of a complex pattern of dust filaments
tracing a two-arm spiral and penetrating into the center of the
galaxy. Similar dust spirals were found by HST imagers in the
centers of many galaxies during the last years (e.g. Martini \&\
Pogge 1999, Regan \&\ Mulchaey 1999, Barth et al. 1995).

There is a region between the nucleus and a spiral arm, at
$r=5\arcsec - 12\arcsec$ to the south of the nucleus, where
H$\alpha$ emission-line profiles are clearly two-component (
Fig.~\ref{fig_prof}\,i). The main line component traces the mean
rotation; the secondary component is redshifted by $\sim 150
\km$, sometimes being as strong as the main one. At the outskirts
of this region the H$\alpha$ profiles often possess a red wing,
so at the residual velocity map the whole area located between
bright HII regions is distinguished by strong non-circular motion
reaching $50 - 70 \km$. The diameter of the anomalous area is
$10\arcsec - 15\arcsec$ ($1 - 1.5$ kpc for the adopted galaxy
distance of 18 Mpc). As this area is located at the galactic
minor axis, the velocity residuals will be caused by radial gas
motions if they are confined to the plane of the galactic disk.

\subsection{NGC 3583}

This galaxy is a member of the pair of galaxies shared with
NGC~3577 (their separation is 5\arcmin), but there is no clear
evidence of interaction. A small elliptical companion may be
found at a distance of about 1\arcmin ~(see the note in RC2
catalog). The continuum image of the galaxy reveals the presence
of a bar which can be traced up to $r\approx 30\arcsec$ and is
aligned with $PA = 80\degr$; it also contains a lot of HII regions
(Figs.~\ref{ngc3583}\,a and \ref{ngc3583}\,b). Isovelocities
close to the bar have a typical S-like shape
(Fig.~\ref{ngc3583}\,c), the kinematic major axis turns here by
more than $20\degr$, which is evidence of a significant
contribution of non-circular motion into the line-of-sight
velocity field (Fig.~ \ref{ngc3583}\,d). Indeed, within the bar,
in the radius range of $2\arcsec - 18\arcsec$, the line-of-sight
velocities differ from the mean rotation velocity field by $\pm
(50 - 80) \km$. The `blue'' and ``red'' residual velocity spots
are located symmetrically on both sides of the center, almost
along the minor axis of the galactic disk, which enables us to
propose a radial gas flow due to the bar. The presence of a broad
stellar H$\alpha$ absorption line in the circumnuclear region of
this galaxy introduces some uncertainty into the line-of-sight
velocity estimates, but in comparison to other analogous cases
(IC~1525, NGC~4414, etc), this effect is not as severe.

\subsection{NGC 3893}

The galaxy forms a pair with NGC~3896 without any sign of tidal
interaction (their separation is $3\farcm 9$). The rotation curve
rises all along the radius, having no clear maximum (Fig.
~\ref{ngc3893}\,d). The turn of the kinematic major axis at
$r<20\arcsec$ is related to the velocity field perturbation
caused by a small bar aligned at $PA\approx 20\degr$, which is
barely seen in the continuum image (Fig.~\ref{ngc3893}\,a). The
variations of $PA_0$ at $r=20\arcsec - 50\arcsec$ may be a spiral
arm effect. At the radii of $37\arcsec - 45\arcsec$ in the
direction of $PA=140\degr$, at the inner edge of the spiral arm,
the H$\alpha$ emission-line profiles have a double-peaked
structure ( see Fig.~\ref{fig_prof}\, j), the secondary
(non-circular) component being redshifted by $+50 \km$. The
region where the secondary velocity component appears has a
diameter of $8\arcsec - 9\arcsec$ (about 500 pc for the adopted
distance of 13 Mpc), is located between bright HII regions, and
is related to local dynamical processes in the gaseous disk
rather than to the spiral arms.

The orientation angles of this galaxy were also estimated from
the $H\alpha$ velocity field by Garrido et al. (2002), by
minimizing the dispersion of the points along the rotation curve.
The value found by them, $PA_0=347\degr$, is close to the
$345\degr \pm 2\degr$ accepted in this paper, but the difference
of inclination angles ($30\degr$ versus $45\degr \pm 5\degr$) is
significant. Note however that the accuracy of measurements of
$i$ given by Garrido et al. (2002) is rather low, $\pm (5\degr -
10\degr)$.

\subsection{NGC 4100}

This galaxy is a member of the Ursa Major cluster (Odenwald 1986),
although it has no noticeable neighbors -- the distance to the
nearest luminous cluster galaxy is no less than 210 kpc (see the
map of the cluster in Tully et al. 1996). Both CCD images and
surface brightness profiles reveal a central ``lens'' between the
radii of 15\arcsec\ and 70\arcsec\, containing a pair of tightly
wound spiral arms. Beyond the ``lens'', the surface brightness
falls abruptly, although some interesting details -- particularly
two low-contrast symmetric arms -- can still be seen in the outer
parts of the disk.

Our IFP data covers only the region of the central ``lens''; here
the H$\alpha$ emission is confined to the spiral arms and to the
very bright circumnuclear region, $r<12\arcsec$
(Fig.~\ref{ngc4100}\,b). According to Pogge (1987), this
circumnuclear region concentrates more than 20\%\ of the total
H$\alpha$ emission of the galaxy. The emission-line intensity
ratios in the nuclear spectrum are typical for HII regions (Ho et
al. 1997), hence a high H$\alpha$ brightness of the nucleus may
be attributed to intense star formation burst, distinguished in
the background of the more quiescent galactic disk.

A kinematic study of this galaxy's ionized gas with a long-slit
spectrograph (Afanasiev et al. 1988b, 1992) has shown that the
circumnuclear region of NGC~4100 is dynamically decoupled from the
rest of the disk: its rotation curve demonstrates two local
extrema at $r=\pm 5\arcsec$, with a rotation velocity gradient of
about $300 \km {\rm kpc}^{-1}$ between them, which may be
evidence of a highly concentrated bulge (Afanasiev et al. 1988b).
Above all, Afanasiev et al. (1992) noted that the kinematic major
axis in the central region, $r<5\arcsec - 6\arcsec$, was turned
by about $25\degr$ with respect to the outer isophote
orientation. The authors treated this fact as a result of minibar
influence in the center of this galaxy.

Our analysis of the velocity field of t he ionized gas in NGC~4100
(Fig.~\ref{ngc4100}\,c) confirms the dynamical decoupling of the
circumnuclear region, although the local maximum of the rotation
curve (Fig.~ \ref{ngc4100}\,d) is not as prominent as it appeared
followed from Afanasiev et al. data. Besides the turn of the
kinematic major axis, from $PA_0 = 350\degr \pm 2 \degr$ for the
outer galaxy to $358\degr \pm 3\degr$ for the inner one
(Fig.~\ref{ngc4100}\,d), the inclination determined from the
kinematics in the tilted-rings model changes abruptly from $i =
72 \degr \pm 2 \degr$ for the outer galaxy to $i = 60 \degr \pm
5\degr$ for the inner one. As it was noted in the third section
of this paper, the change of $i$ in a tilted-rings model may be
just a reflection of the ordered non-circular coplanar motion of
a gas. In general, presence of a bar may easily imitate the
inclination variations in a line-of-sight velocity field (Moiseev
\&\ Mustsevoi 2000). But in this particular galaxy, the
photometric major axis turns in the same direction  as the
kinematic one, so we cannot exclude the presence of an inclined
gas disk in the circumnuclear region of NGC~4100 (Zasov \&\
Moiseev 1999). The radius of this disk is less than 11\arcsec\ -
12\arcsec\ (0.8 kpc for the adopted distance of 16 Mpc).
Depending on which side of the circumnuclear disk is nearest to
us, the inclination of the central disk to the global galactic
plane must be either $25\degr$ or $55\degr$.

Our rotation curve for NGC~4100 (Fig.~\ref{ngc4100}\,d)
demonstrates a monotonic rise of the rotation velocity up to the
spatial limits of the measurements, which is consistent with HI
observations of Verheijen (1996), demonstrating that the rotation
velocity rises up to $r=90\arcsec$; in the very outer part of the
galaxy the rotation velocity reaches a plateau.

\subsection{NGC 4136}

According to Allsopp (1979), the galaxy is a possible member of
the CVnII group. In the central part ($r<15\arcsec$) there is a
small bar weakly recognizable in our continuum image (Fig.~
\ref{ngc4136}\,a), but clearly seen on the CCD images given in
The Digital Atlas of Nearby Galaxies by Frei et al. (1996). The
galaxy is seen almost face-on, and as a result, the parameters $i$
and $PA_0$ are poorly determined (see the dispersion of the
estimates in Table~\ref{orpar}). For this reason, the absolute
value of the rotation velocity is rather uncertain. The
``troughs'' at the rotation velocity (Fig.~\ref{ngc4136}\,d) may
be just artifacts. Formally, this uncertainty is expressed by
large error bars in Fig.~\ref{ngc4136}. Under such a disk
orientation, the line-of-sight velocity deviations from the mean
rotation in most cases may be related mostly to gas motion along
$z$-coordinate direction.

At the radii of $4\arcsec - 15\arcsec$ the H$\alpha$ emission is
either negligible, or strongly contaminated by the absorption
line due to bulge stars; this effect cannot be properly taken
into account because of the IFP interference order overlapping.

\subsection{NGC 4414}

According to Elmegreen \&\ Elmegreen (1987), NGC~4414 is a
flocculent galaxy; our continuum image (Fig.~\ref{ngc4414}\,a)
does not show any traces of a spiral pattern either. However,
observations in the NIR bands reveal the presence of several
extended segments of spiral arms populated by old stars (Thornley
1996). In the circumnuclear region of NGC~4414 we see a narrow
H$\alpha$ emission line inside the broad absorption produced by a
contribution of bulge stars; within several arcseconds from the
galactic center only the [NII] emission line from the neighboring
order of interference is seen (Fig.~\ref{fig_prof}\, k). But
unlike other similar cases (IC~1525, NGC~615, etc), for this
galaxy we have been able to construct a velocity field of the
ionized gas in the central region by subtracting the absorption
line superposed and by applying a Gaussian analysis to the
H$\alpha$ and [NII] emission-line profiles. As a result, only a
small area with a size of 7\arcsec\ to the south-east of the
nucleus remained uncovered by our measurements
(Fig.~\ref{ngc4414}\,c). Our orientation parameters for NGC~4414
($PA=156\degr$, $i=56\degr$) are very close to those obtained by
Thornley \&\ Mundy (1997) from the analysis of the HI and CO
velocity fields ($PA=160\degr$, $i=60\degr$). The HI velocity
field measured in this work extends beyond three optical radii of
the galaxy and demonstrates a strong asymmetry of the outer disk
at $r>200\arcsec$, although no optical satellite is found within
the nearest outskirts of NGC~4414 (Thornley \&\ Mundy 1997).

The turn of the kinematic major axis and large error bars of the
mean rotation velocity estimates at $r<9\arcsec$
(Fig.~\ref{ngc4414}\,d) may be caused by non-circular gas motion
in the minibar area. The presence of a minibar in NGC~4414 was
first suspected by Thornley \&\ Mundy (1997) from isovelocity
twisting seen in the HI velocity field, though the spatial
resolution of the HI observations was much worse than our
resolution of the ionized gas observations.

\subsection{NGC 4814}

The main fraction of the H$\alpha$ emission in this galaxy is
confined to the inner disk having a radius of 50\arcsec; beyond
this radius the emission regions are concentrated in a pair of
spiral arms (Fig.~13b). A mean rotation curve
(Fig.~\ref{ngc4814}\,d) agrees well with the rotation curve of
the inner region, $r<30\arcsec$, we obtained from the earlier
long-slit observations at the 6m telescope (Zasov \&\ Sil'chenko
1987; Afanasiev et al. 1988b). A turn of the kinematic major axis
at $r<15\arcsec$ may be related to a small bar, but it is not
accompanied by the significant continuum isophote turn. Afanasiev
et al. (1988b) reported non-circular gas motion to the west of the
nucleus, at $r\approx 20\arcsec$. After subtracting the mean
rotation field from our velocity data, we have found that the
residual velocities in this region are about $-40 \km$, evidently
produced by an effect of the spiral pattern on the kinematics of
the gaseous disk.

\subsection{NGC~5371}

This galaxy possesses a modest bar of about $20\arcsec -30\arcsec$
length aligned with the minor axis of the galactic disk. The bar
is barely seen on our continuum image (Fig.~\ref{ngc5371}\,a), but
reveals itself rather well on the other optical images (e.g.
Gonzalez et al. 1996). Unfortunately the IFP observations of this
galaxy were made under poor seeing conditions during a
non-photometric night. Due to this, we present velocity
measurements only for the brightest emission regions in the
spiral arms (Fig.~\ref{ngc5371}\,c). In the circumnuclear region
the H$\alpha$ emission line is strongly broadened, and the
intense [NII] emission line from the neighboring interference
order is overlaid (the [NII] emission is abnormally strong
because NGC~5371 is classified as a LINER (Rush et al. 1993)). We
have not succeeded in separating the contributions of the
overlapping broad emission lines into the integrated profile.
Perhaps due to this uncertainty our measurements of the
line-of-sight gas velocities in the circumnuclear region do not
reveal the local rotation velocity maximum at $r=10\arcsec -
15\arcsec$ reported by Zasov \&\ Sil'chenko (1987). It is worth
noting that the inclination angle $i$ we obtained from the
velocity field, $i=47\degr \pm 8\degr $, is somewhat larger than
$i=34\degr$, adopted by Wevers et al. (1986).

\subsection{NGC 6643}

According to the classification of Elmegreen \&\ Elmegreen (1987),
the galaxy has a 5th arm class, and therefore is almost lacking an
ordered spiral structure. Indeed, both our continuum and H$\alpha$
images (Fig.~\ref{ngc6643}\,a and \ref{ngc6643}\,b) present a
flocculent-like galaxy. However, recent NIR observations by
Elmegreen et al. (1999) reveal the presence of a two-arm spiral
pattern in NGC~6643 related to an old stellar population and
implying a dynamical origin. Afanasiev et al. (1988b) noted
large-scale non-circular gas motion in the central part of the
galactic disk of this galaxy. Our analysis of the velocity field
shows that line-of-sight velocity deviations from the pure
rotation field do exist, but they do not exceed $40-50 \km$ over
the whole disk of the galaxy. These motions are probably related
to the spiral structure; but an exact diagnosis will be possible
only after a detailed Fourier analysis of the velocity field,
which is under way. However there are a few regions in the galaxy
where the H$\alpha$ emission-line profile cannot be fitted by a
single Gaussian function. For example, at $r=12\arcsec$ to the
north-west of the nucleus, between bright HII regions, the
H$\alpha$ profiles demonstrate a broad red wing that may be
evidence of the presence of the secondary ``non-circular''
velocity component of $100-200 \km$ in projection on the line of
sight (see Fig.~\ref{fig_prof}\, ,l). The size of this anomalous
velocity area is $5\arcsec \times 10\arcsec$ (about 1 kpc, under
the adopted distance of 19 Mpc). The effect of this local
velocity anomaly on the whole velocity field is rather strong: it
is responsible for a turn of the kinematic major axis observed in
the radius range of $r=3\arcsec - 12\arcsec$
(Fig.~\ref{ngc6643}\,d). If we mask this region when analyzing
the velocity field, the position angle of the kinematic major
axis in the central region becomes about $40\degr$, in good
agreement with the orientation of the line of nodes of the outer
disk (Table~\ref{orpar}). We should also note that the inclination
$i\approx 60\degr$ found in this work is less than $i=68\degr$,
used by Afanasiev et al. (1988b) but agrees with $i=59\degr$ of
Bottinelli et al. (1984) obtained from photometric data.

\section{Discussion} \label{sec5}

Our sample of galaxies is rather widely representative, because it
includes objects with a variety of kinematic properties of gas and
with a wide range of structural features. The rotation curves in
all the cases rise steeply in the inner parts within $r \approx
10\arcsec - 20\arcsec$ (in most cases inside 1 - 3 kpc from the
center) due to the combined potentials of the bulge and the disk;
further from the center the shape of curves looks different: the
curves have either a plateau with some local velocity variations,
or pass local maxima (as in the cases of NGC 23, 615, 1134, 3583,
5371). The curves of rotation in IC 1525, NGC 972, 1084, 3893,
4100, 4136, 6643 rise to the limit of the observed part of the
galaxies (which, being less than the optical sizes of the
galaxies, nevertheless contains the brightest part of the spiral
structure). The opposite tendency is observed in NGC 2964; its
rotation curve falls beyond the region of bright spiral arms.

The observed large-scale non-circular motions are usually
connected with a spiral density wave or a bar. In particular,
these motions result in the artificial $PA_0$ and $i$ radial
variations in a tilted-rings model, even when their real values
remain constant. Systematic perturbations of the line-of-sight
velocities in the region of spiral arms are most clearly seen in
NGC 23, 4814, and 6643 (see, respectively, Figs. \ref{ngc23}\,d,
\ref{ngc4814}\,d and \ref{ngc6643}\,d), although in other galaxies
they also appear.

In the presence of a bar (which can be considered another
manifestation of a density wave) or some other type of triaxial
structure (triaxial bulge, oval lense, etc) gas clouds follow
elongated orbits which results in an observed turn of the
kinematic major axis and leads to a well-known S-shaped
distortion of isovelocities (see e.g. Chevalier \&\ Furenlid
1978). Besides that, in such cases a secondary component of the
emission line is often observed; as a rule, it traces radial gas
flows along the bar. Indeed, these peculiarities are observed in
our galaxies, but curiously, the velocity anomalies correlate
weakly with the relative size and contrast of a bar. There are
three galaxies in our sample classified as SB in both NED and LEDA
catalogs: IC~1525, NGC~23, and NGC~3583. In the former two we
observe only a slight hint of bar-related anomalies, and only in
NGC~3583 the expected distortion is clearly observed. In general,
among the galaxies with the kinematic signatures of a bar there
are the galaxies with the known photometric bars (SAB and SB,
according to NED: IC~1525, NGC~23, NGC~2964, NGC~3583) as well as
the galaxies previously classified as SA (NGC~615, NGC~972,
NGC~4414, NGC~4814). However it is well known that low-contrast
optical, or only infrared, triaxial structures may exist in
galaxies classified as SA (NGC~972, NGC~615). Indeed, in NGC~972
the bar has been detected from the NIR surface photometry (Mayya
et al. 1998), and in NGC~615 an oval inner disk has been found
from the optical photometric data (Sil'chenko et al. 2001). We
think also that the local regions with large residual gas
velocities, situated symmetrically on opposite sides of the
center and characterized by an enhanced [NII] emission, which are
found in NGC 1084 and NGC 1134, may be connected to the front of a
bar-induced shock wave, although optical bars are not clearly
seen here.

Besides the large-scale deviations from circular rotation related
to a bar or spiral arms, significant small-scale velocity
anomalies are also observed in all galaxies we studied. They may
be divided into three categories.

The first type refers to the non-circular motions of different
natures in circumnuclear regions (within ten arcseconds from the
center). Wide wings or significant shifts of spectral lines may
be evidence for radial gas motion as a result of nuclear activity
or an active star formation in the nuclei (IC~1525, NGC~23,
NGC~972, NGC~1084). One should bear in mind, however, that
similar manifestations (especially from observations with a
moderate spatial resolution) may be caused by nuclear mini-bars
(NGC~615, NGC~972, NGC~2964, NGC~4414) or by an inclined nuclear
disk (NGC~4100). More detailed analysis is needed to distinguish
between these cases (see discussion in Zasov \&\ Moiseev 1999).

The second type of local non-circular motion is observed in the
bright HII regions of a disk. Among our galaxies such motions are
definitely observed only in the most actively star-forming galaxy
NGC~972. The velocity distortions are not too high (about $20
\km$). This type of motion is most certainly caused by gas
outflow as a result of heating and blowing by massive stars in
the region of a local star formation burst. More energetic
motions of this type are usually observed in star-forming dwarf
galaxies (radial expansion with amplitudes of ten to a hundred
$\km$, see for example Pustilnik et al. 2001).

The third type of local non-circular motion is the most
enigmatic. It takes place far from the nuclei, in the regions
with a typical diameter of about 0.3-1.5 kpc which exceeds the
size of bright regions usually covered by a single burst of star
formation. Abnormally moving gas is revealed by its Doppler
velocities, which differ from the velocities of the surrounding
disk or from the velocities expected for circular rotation at a
given point. The velocity differences reach some tens of $\km$,
but sometimes they exceed $100 \km$. Such regions are found in
IC~1525, NGC~1084, NGC~2964, NGC~3893, and NGC~6643. These
velocity anomalies are often accompanied by an appearance of
secondary emission-line components which indicate the existence of
at least two independent subsystems of ionized-gas clouds
possessing normal and abnormal motions on the line of sight.
Sometimes the presence of a ``non-circular'' velocity component
is seen in the wide asymmetric wings of the emission line
profiles. In IC~1525, NGC~1084, and partly in NGC~1134, the
non-circular motions are accompanied by an increase in the
[NII]/H$\alpha$ ratio that argues in favor of a shock excitation
mechanism (see the detailed discussion in Moiseev 2000).

The regions of these abnormal velocities are usually observed off
the sites of active star formation, although the bright HII
regions are often seen near them. It seems as if the gas with
abnormal velocity is usually located between the complexes of
HII. Nevertheless it does not mean that the ``abnormal'' gas
definitely flows around normally rotating islands of star
formation: indeed, the extended regions of this gas may cover
bright HII regions in projection, which makes the emission of
abnormally moving gas non-observable against the bright
background.

These regions may be related to powerful ``galactic fountains'',
where gas thrown out of star formation sites along the $z$
coordinate returns, falling toward the galactic disk (see for
example Breitschwerdt \&\ Komossa 2000, de Alvis 2000). We
observe this gas more easily when it is projected onto the areas
of low gas emissivity. The other event which may account for the
observed gas motion is the interaction of low density gas of a
galactic disk with high velocity clouds accreting onto the
galaxy. In this case, the appearance of star formation sites in
this region is a consequence, not the cause, of the abnormal
velocities of the gas. Further studies may distinguish between
these possibilities.

\section{Conclusions}

The main results of this work may be summarized as follows:
\begin{itemize}

\item{The presence of regular noncircular motion of gas related to
spiral density waves may introduce systematic errors in the
evaluation of the orientation parameters $i$ and $PA_0$ of
galactic disks. In particular, it affects the commonly used
tilted-rings model. The systematic errors remain even if the
parameters found for individual rings in this model are averaged.
To avoid this effect, we force the orientation parameters to be
constant all over the disk within some radius. Still, the
tilted-rings model remains a useful diagnostic method of revealing
non-coplanar or strongly non-axisymmetric gas motion at some
distance from the center (in circumnuclear regions or the disk
outskirts).}

\item {For the sample of fifteen observed galaxies the parameters $i$ and
$PA_0$ were estimated using both the kinematic and photometric
methods we describe in this work; the latter is based on the
analysis of the second Fourier harmonic of azimuthal brightness
distribution in the sky plane. Both kinematic and photometric
methods of disk orientation parameter determination may have
their limitations, especially in galaxies with strong triaxial
(oval or bar) distortions.
 However, the close agreement of the orientation
parameters found by two independent methods for most of our sample
galaxies shows that the effect of systematic non-circular gas
motion does not significantly alter the resulting parameters
obtained from kinematic data for a pure circular rotation model.}

\item{We have obtained the curves of rotation for 15 spiral galaxies,
assuming the gas motion to be circular, and found the confidence
intervals for rotation velocities at a given radii which
characterize possible errors due to ignoring the regular
non-circular motion related to spiral arms and/or a bar.}

\item{Practically all the galaxies possess local non-circular gas motion,
which appear either as residual velocities left after subtraction
of circular velocities from the observed velocity fields, or as
the complex profiles of emission lines, which may be decomposed
into circular and non-circular components. Local anomalies of the
velocity fields on a scale of several hundred pc (up to 1-1.5
kpc) are related to star formation or some other mechanisms,
which are responsible for the non-circular velocities often
exceeding $100 \km$ (see the discussion in the previous section.)}

\item {The observed regular large-scale non-circular motion in the
galaxies have another nature: they are related to bars and/or
spiral arms of galaxies. In principle, the detailed analysis of
these density-wave-induced motions may allow to restore a full 3D
vector velocity field of the gas in galactic disks (see the
Introduction). This will be done as a future work.}

\end{itemize}

\begin{acknowledgements}

The authors express their thanks to J.Boulesteix and S.Drabek for
the software development used in the data processing and
S.D.Ryder for the possibility to use IR images of two galaxies.
We especially thank the referee, Dr.Y.Copin for his fruitful comments
and remarks. We also appreciate the organizers of the HYPERCAT
Database which was used in this paper. This research is partially
based on data from the ING Archive and used the NASA/IPAC
Extragalactic Database (NED) operated by JPL under contract with
NASA. The authors also thank Max Fridman for improving the
quality of the text.

This work was performed under partial financial support from RFBR
grant N 02-02-16878, grant ``Leading Scientific Schools'' N
00-15-96528, and the contracts with Ministry of Industry, Science
and Technology \N \N 40.022.1.1.1101 and 40.020.1.1.1167.

\end{acknowledgements}

\newpage

\section{Appendix. Images and gas velocities of the individual
galaxies.}

See the text for details

\begin{figure*}
\centering
\caption[]{The same as
fig.\ref{ic1525} for NGC~23. } \label{ngc23}
\end{figure*}

\newpage

\begin{figure*}
\centering
\caption[]{The same as
fig.\ref{ic1525} for NGC~615. } \label{ngc615}
\end{figure*}

\newpage

\begin{figure*}
\centering
\caption[]{The same as
fig.\ref{ic1525} for NGC~972.} \label{ngc972}
\end{figure*}

\newpage

\begin{figure*}
\centering
\caption[]{The same as
fig.\ref{ic1525} for NGC~1084.} \label{ngc1084}
\end{figure*}

\newpage

\begin{figure*}
\centering
\caption[]{The same as
fig.\ref{ic1525} for NGC~1134. } \label{ngc1134}
\end{figure*}

\newpage

\begin{figure*}
\centering
\caption[]{The same as
fig.\ref{ic1525} for NGC~2964.} \label{ngc2964}
\end{figure*}

\newpage

\begin{figure*}
\centering
\caption[]{The same as
fig.\ref{ic1525} for NGC~3583. There is a ghost-image in the
bottom-left corner of the monochromatic and continuum images.  }
\label{ngc3583}
\end{figure*}

\newpage

\begin{figure*}
\centering
\caption[]{The same as
fig.\ref{ic1525} for NGC~3893.} \label{ngc3893}
\end{figure*}

\newpage

\begin{figure*}
\centering
\caption[]{The same as
fig.\ref{ic1525} for NGC~4100.} \label{ngc4100}
\end{figure*}

\newpage

\begin{figure*}
\centering
\caption[]{The same as
fig.\ref{ic1525} for NGC~4136.} \label{ngc4136}
\end{figure*}

\newpage

\begin{figure*}
\centering
\caption[]{The same as
fig.\ref{ic1525} for NGC~4414.} \label{ngc4414}
\end{figure*}

\newpage

\begin{figure*}
\centering
\caption[]{The same as
fig.\ref{ic1525} for NGC~4814.} \label{ngc4814}
\end{figure*}

\newpage

\begin{figure*}
\centering
\caption[]{The same as
fig.\ref{ic1525} for NGC~5371.} \label{ngc5371}
\end{figure*}

\newpage

\begin{figure*}
\centering
\caption[]{The same as
fig.\ref{ic1525} for NGC~6643.} \label{ngc6643}
\end{figure*}

\end{document}